\documentclass[12pt]{article}
\usepackage{amsmath,amssymb,latexsym,epsfig,cite}

\begin{document}

\begin{flushright}
LNF - 10 / 19(P)
\end{flushright}

\begin{center}
{\Large \bf Operational methods and Lorentz-type \\ equations of motion}
\end{center}

\begin{center}
D. Babusci$^1$, G. Dattoli$^2$, E. Sabia$^3$\\[2ex]
{ $^{1}$\it INFN - Laboratori Nazionali di Frascati, v.le E. Fermi, 40, 00044 
Frascati (Roma), Italy}\\
{ $^{2}$\it ENEA -  Centro Ricerche Frascati, v.le E. Fermi, 45, 00044,  
Frascati (Roma), Italy}\\
{ $^{3}$\it ENEA -  Centro Ricerche Portici, P.le E. Fermi, 80055, 
Portici (Napoli), Italy}
\end{center}

\begin{abstract}
We propose an operational method for the solution of differential equations involving vector products. 
The technique we propose is based on the use of the evolution operator, defined in such a way that the 
wealth of techniques developed within the context of quantum mechanics can also be exploited for classical 
problems. We discuss the application of the method to the solution of the Lorentz-type equations.
\end{abstract}

\section{Introduction} \label{sec:intro}
Operational methods provide powerful techniques to solve problems both in classical and quantum mechanics. 
The distinctive feature of these tools is their versatility and the possibility of exploiting them in absolutely different 
contexts, from the time dependent Schr\"{o}dinger problems to the charged beam transport in accelerators. 
Differential equations have been the primary motivation for the introduction of these techniques. Lie introduced the 
algebras bearing his name within the framework of a program aimed to clarify the reasons why only restricted 
families of ordinary differential equations (ODE) can be solved by quadrature. Operational methods  have 
become ``popular'' in applied science for their wide flexibility and have stimulated the development of new 
computer languages, useful for symbolic manipulation.

In this paper we go back to the solution of some differential equations involving vector products and we will further 
develop the point of view suggested in Ref. \cite{dattoli}, by presenting an analysis which includes a variety of 
problems often encountered in applications, including classical time ordering techniques, which are not widespread 
known as they should.

We will start our analysis with the following Cauchy problem
\begin{equation}\label{eq:cauchy}
\frac{\mathrm{d}}{\mathrm{d}t}\,\vec{S} \,=\, \vec{T}\,\times\,\vec{S}\;,\qquad\qquad\vec{S}|_{t = 0}\,=\,\vec{S}_0\;,
\end{equation}
almost ubiquitous in physics, from classical mechanics to nuclear magnetic resonance. Although this equation 
can always be written in a matrix form, we will develop our considerations using the vector notation and the 
properties of the vector product because they are more concise and more insightful from the physical point of 
view. We will assume, for the moment, that the torque vector $\vec{T}$ is not explicitly time dependent, 
so that a straightforward application of the evolution operator formalism yields
\begin{equation}
\vec{S} (t)\,=\, \hat{U}(t)\,\vec{S}_0\;, \qquad\qquad \hat{U}(t) \,=\,\mathrm{e}^{t\,\hat{T}}\, 
\qquad (\hat{U}(0) \,=\,\hat{1} )\;, 
\end{equation}
where $\hat{U}(t)$ is the evolution operator defined in terms of the operator $\hat{T}$, whose properties 
will be specified below. The series expansion of the exponential provides the following solution:
\begin{equation}
\vec{S} (t)\,=\, \sum_{n = 0}^\infty \,\frac{t^n}{n!}\,\hat{T}^n \,\vec{S}_0 
\end{equation}
where $\hat{T}$, called \emph{vector evolution operator} (VOP), satisfies the following identities:
\begin{equation}\label{eq:opern}
\hat{T}^0 \,\vec{S}_0 \,=\, \vec{S}_0\; 
\quad\cdots\quad \hat{T}^n \,\vec{S}_0 \,=\, \underbrace{\vec{T}\,\times\,(\vec{T}\,\times\,\cdots\,
(\vec{T}\,\times}_{n - 1\; \text{times}}\,(\vec{T}\,\times\,\bar{S}_0))\,\cdots)\;.
\end{equation}
The use of the cyclical properties of the vector product leads to the following closed form for $\vec{S}$ \cite{dattoli} 
($T = |\vec{T}|$)
\begin{equation}\label{eq:svect}
\vec{S} (t)\,=\, \cos (T\,t)\,\vec{S}_0 \,+\, \sin (T\,t)\,(\vec{n} \times \vec{S}_0) \,+\, \left[1 - \cos (T\,t)\right]\, 
(\vec{n} \cdot \vec{S}_0)\,\vec{n}\, \qquad (\vec{n} =  \vec{T}/T)\;.
\end{equation}
This solution has an almost natural geometrical interpretation, which is recognized as a Rodrigues rotation (R. r.) 
\cite{rodrig}. 

The solution of the non-homogeneous version of eq. (\ref{eq:cauchy})
\begin{equation}\label{eq:caunoh}
\frac{\mathrm{d}}{\mathrm{d}t}\,\vec{S} \,=\, \vec{T}\,\times\,\bar{S} \,+\, \vec{N}\;,
\qquad\qquad\vec{S}|_{t = 0}\,=\,\vec{S}_0\;,
\end{equation}
is given by
\begin{equation}
\label{eq:solcnh}
\vec{S} (t)\,=\, \hat{U} (t)\,\left(\vec{S}_0 \,+\, \int_0^t \,\mathrm{d}t^{\prime}\,
\hat{U}^{-1} (t^{\prime})\,\vec{N}\right)\,,
\end{equation}
which is the solution of the first ODE with the evolution operator in place of the integration factor. 

The results we have obtained so far are a more concise form for the well known solutions of equations 
of the type (\ref{eq:caunoh}),  often encountered in the study of problems involving the Coriolis \cite{goldst} 
and Lorentz \cite{jacks} forces. In the next sections we will discuss specific applications of the outlined 
formalism and we will see how it may provide further progress when applied to actual physical problems, 
including time-dependent or spatially non-homogeneous fields.

\section{The Lorentz equation of motion}\label{sec:lore}
\begin{figure}[htb]
\centering
\includegraphics[height=9cm]{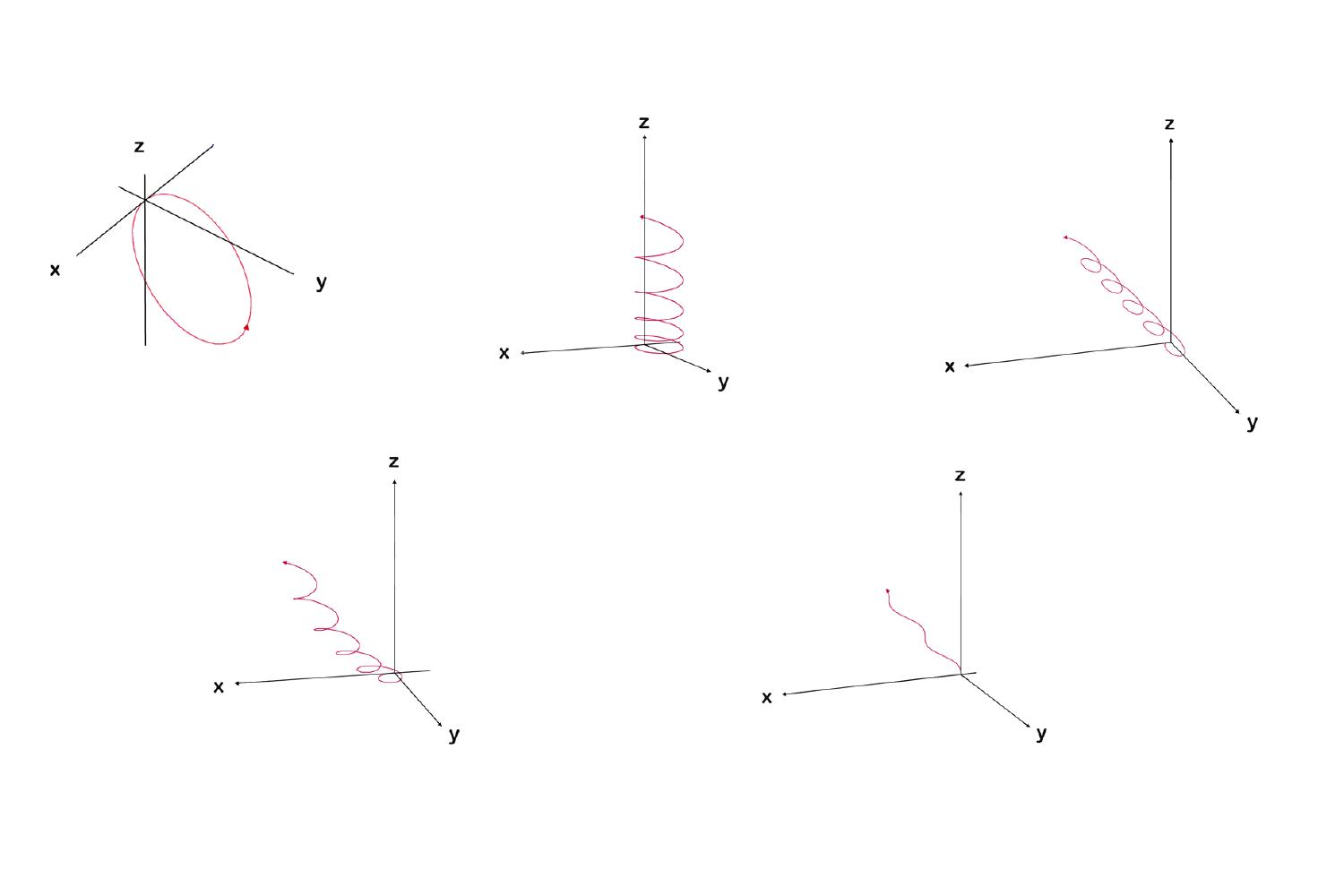}
\caption{Trajectory followed by an electron for different configurations of electric and magnetic 
fields and initial velocity: 
(upper panel, left) $\vec{v}_0 = (v_0, 0, 0), \vec{B} = (0, B/\sqrt{2}, B/\sqrt{2}), \vec{E} = 0$
(upper panel, middle) $\vec{v}_0 = (v_0/\sqrt{2}, 0, v_0/\sqrt{2}), \vec{B} = (0, 0, B), \vec{E} = (0, 0, -E)$
(upper panel, right) $\vec{v}_0 = (v_0/\sqrt{2}, 0, v_0/\sqrt{2}), \vec{B} = (0, 0, B), \vec{E} = (0, 0, -E)$
(lower panel, left) $\vec{v}_0 = (v_0/\sqrt{2}, 0, v_0/\sqrt{2}), \vec{B} = (0, 0, B), \vec{E} = (0, E/\sqrt{2}, E/\sqrt{2})$ 
(lower panel, right)  $\vec{v}_0 = (v_0/\sqrt{2}, 0, v_0/\sqrt{2}), \vec{B} = (0, 0, B), \vec{E} = (0, E, 0)$.}
\label{fig:traj} 
\end{figure}

The non-relativistic dynamics of a particle with mass $m$ and charge $e$ under the combined influence of static 
electric and magnetic fields, is ruled by the Hamiltonian ($c = 1$):
\begin{equation}\label{eq:hamil}
H \,=\, \frac{1}{2 m}\,(\vec{p} - e \vec{A})^2 \,+\, e\,\Phi\;,
\end{equation}
where $\vec{p} $ is the canonical momentum. In the \textit{static symmetric} gauge, the vector and scalar potentials 
are given by \cite{JH}
\begin{equation}\label{eq:vspot}
\vec{A} \,=\, \frac{1}{2}\,\vec{B} \times \vec{r}\;, \qquad \Phi \,=\, -\vec{E} \cdot \vec{r}\;. 
\end{equation}
The equation of motion for the mechanical momentum 
\begin{equation}
\vec{\pi} \, =\, m\,\vec{v}\,=\, \vec{p} \,-\, \frac{e}{2}\, \vec{B} \times \vec{r}\;,\nonumber
\end{equation}
derived from eq. (\ref{eq:hamil}), is the Lorentz equation:
\begin{equation}\label{eq:lorentz}
\frac{\mathrm{d}}{\mathrm{d}t}\vec{v} \,=\,-\,\vec{\Omega} \times \vec{v} \,+\, \vec{Q}\;,
\end{equation}
where we have introduced the following vectors
\begin{equation}
\vec{\Omega} \,=\, \frac{e}{m}\,\vec{B}\;, \qquad \qquad \vec{Q} \,=\, \frac{e}{m}\,\vec{E}\;.
\nonumber
\end{equation}
This equation has the same form of eqs. (\ref{eq:caunoh}) ($\vec{T} = -\,\vec{\Omega}$, $\vec{N} = \vec{Q}$) and 
its solution is given by eq. (\ref{eq:solcnh}) with $\hat{U} (t) =  \mathrm{e}^{- t\,\hat{\Omega}}$. Therefore, we get 
($\vec{n} = \vec{B}/B$):
\begin{equation}\label{eq:veloc}
\vec{v} (t) \,=\, \cos (\omega_c t)\,\vec{v}_0 \,+\, \frac{\sin (\omega_c t)}{\omega_c}\,\vec{Q} \,+\, 
(\vec{n} \cdot \vec{\ell}\,)\,\vec{n} \,-\, \vec{n} \times \vec{m}\;, 
\end{equation}
where $\omega_c = \Omega = eB/m$ is the cyclotron frequency, and we put 
\begin{equation}
\vec{\ell} \,=\, \left[1 - \cos (\omega_c t)\right]\, \vec{v}_0 \,+\, \left[t \,-\, \frac{\sin (\omega_c t)}{\omega_c}\right]\,
\vec{Q}\;, \qquad \vec{m} \,=\, \frac{1}{\omega_c}\,\frac{\mathrm{d}}{\mathrm{d}t}\vec{ \ell}(t)\;. \nonumber
\end{equation}
A further integration, with respect to the time, yields the position vector, which reads:
\begin{equation}\label{eq:posiz}
\vec{r} (t)\,=\, \vec{r}_0 \,+\, \frac{\sin (\omega_c t)}{\omega_c}\,\vec{v}_0 \,+\, 
\frac{1 - \cos (\omega_c t)}{\omega_c^2}\,\vec{Q} \,+\, 
(\vec{n} \cdot \vec{o}\,)\,\vec{n} \,-\, \frac{1}{\omega_c}\, \vec{n} \times \vec{\ell}\;,
\end{equation}
where
\begin{equation}
\vec{o} \,=\, \left[t \,-\, \frac{\sin (\omega_c t)}{\omega_c}\right]\,\vec{v}_0 \,+\, \left[\frac{t^2}{2} \,-\, 
\frac{1 - \cos (\omega_c t)}{\omega_c^2}\right]\,\vec{Q}\;. \nonumber
\end{equation}
(The quantity $r_L = v_0/\omega_c$ is called Larmor radius). The trajectories of an electron for different fields configurations 
are reported in Fig. (\ref{fig:traj}). 

It is interesting to stress the contribution $\vec{n} \times \vec{m}$, that appears when the initial velocity and/or 
the electric field are not parallel to the magnetic field. It is given by:
\begin{equation}\label{eq:vdrift}
\vec{n} \times \vec{m} \,=\, \sin (\omega_c t)\,\frac{\vec{B} \times \vec{v}_0}{B} \,-\,
2 \sin^2 \left(\frac{\omega_c t}{2}\right)\,\vec{v}_d\;, \qquad\qquad \vec{v}_d \,=\, \frac{\vec{E} \times \vec{B}}{B^2}\;,
\end{equation}
where $\vec{v}_d$ is the drift velocity \cite{JH,rothm}. 
In physical terms it can be understood as the component of the 
velocity which allows the balance between electric and magnetic forces in the Lorentz equation of motion. In fact, if 
we decompose the velocity as $\vec{v} = \vec{u} + \vec{V}$, and impose that
\begin{equation}
e\,\vec{B} \times \vec{V} \,=\, e\,\vec{E}\,,
\end{equation}
solving for $\vec{V}$, we recognize that, in the case $\vec{B} \cdot \vec{V} =0$,  this vector coincides with the drift velocity 
vector defined in eq. (\ref{eq:vdrift}). 
It is evident that the presence of a further force acting on the particle will induce 
an analogous drift. By replacing  $\vec{E}$ with $\vec{E} + \vec{E}^\prime$ in the Lorentz equation, we obtain the composed 
drift velocity
\begin{equation}
\label{eq:genvd}
\vec{v}_d \,=\, \frac{\left(\vec{E} \,+\, \vec{E}^\prime\right) \times \vec{B}}{B^2}\;.
\end{equation}
The role played by the drift velocity can be further stressed by noting that it can be associated to the conservation of  
the so called pseudo-momentum vector \cite{JH,rothm,avron}
\begin{equation}
\vec{\Pi} \,=\, m\,\left(\vec{v} - \vec{v}_d\right) \,+\, e\,\left(\vec{B} \times \vec{r} \,-\, t\,\vec{E}\right)\;,
\end{equation}
which is ensured by eqs. (\ref{eq:lorentz}), and has been used as an a posteriori check of the correctness of 
our computations.

\section{Lorentz-type equations and damping}\label{sec:damp}
It is evident that terms of non-electric/magnetic nature\footnote{We mean that the magnetic and/or electric field do not appear 
explicitly in its expression. As shown in the following, genuine non-electric/magnetic effects, like the gravitational term, can 
also be included.} can be added to eq. (\ref{eq:hamil}).  An example is represented by the simplified Drude-like models 
\cite{gropar}, where is introduced a term depending on the velocity and a relaxation time and that may be associated with 
electromagnetic-type interactions. From the mathematical point of view the problem to be treated is the search of the solution 
for the following vector differential equation
\begin{equation}\label{eq:drude}
\frac{\mathrm{d}}{\mathrm{dt}}\vec{v} \,=\, -\,\vec{\Omega} \times \vec{v} \,+\, \vec{Q} \,-\, \frac{1}{\tau}\,\vec{v}\;.
\end{equation}
The presence of a velocity-dependent contribution does not modify the procedure described before and the solution 
is given by eq. (\ref{eq:solcnh}) written for the following evolution operator 
\begin{equation}
\hat{U} (t) \,=\, \exp \left\{-\,t\,\left(\frac{1}{\tau} \,+\, \hat{\Omega}\right)\right\}\,,
\end{equation}
and $\vec{N} = \vec{Q}$. 

The Lorentz equation (\ref{eq:lorentz}) is mathematically equivalent to the one describing the falling of a body under the influence 
of the Coriolis force. It has been stressed, within the framework of the so called gravito-magnetic theories \cite{clark}, that effects 
like the Foucault rotation can be associated to a vector potential of the type reported in eq. (\ref{eq:vspot}) and the relevant analysis 
has been conducted using the perspective of parallel transport and covariant derivative \cite{kugle}. The supposed existence of 
such a potential has given the opportunity of developing further speculations as those associated with a possible observation of the 
Aharonov-Bohm \cite{ahboh} effect involving the Coriolis vector potential. This phenomenon, suggested by Aharanov and Carmi 
\cite{ahcar}, has been observed experimentally using the techniques of neutron interferometry \cite{overh}. 

\begin{figure}[htb]
\centering
\includegraphics[height=6cm]{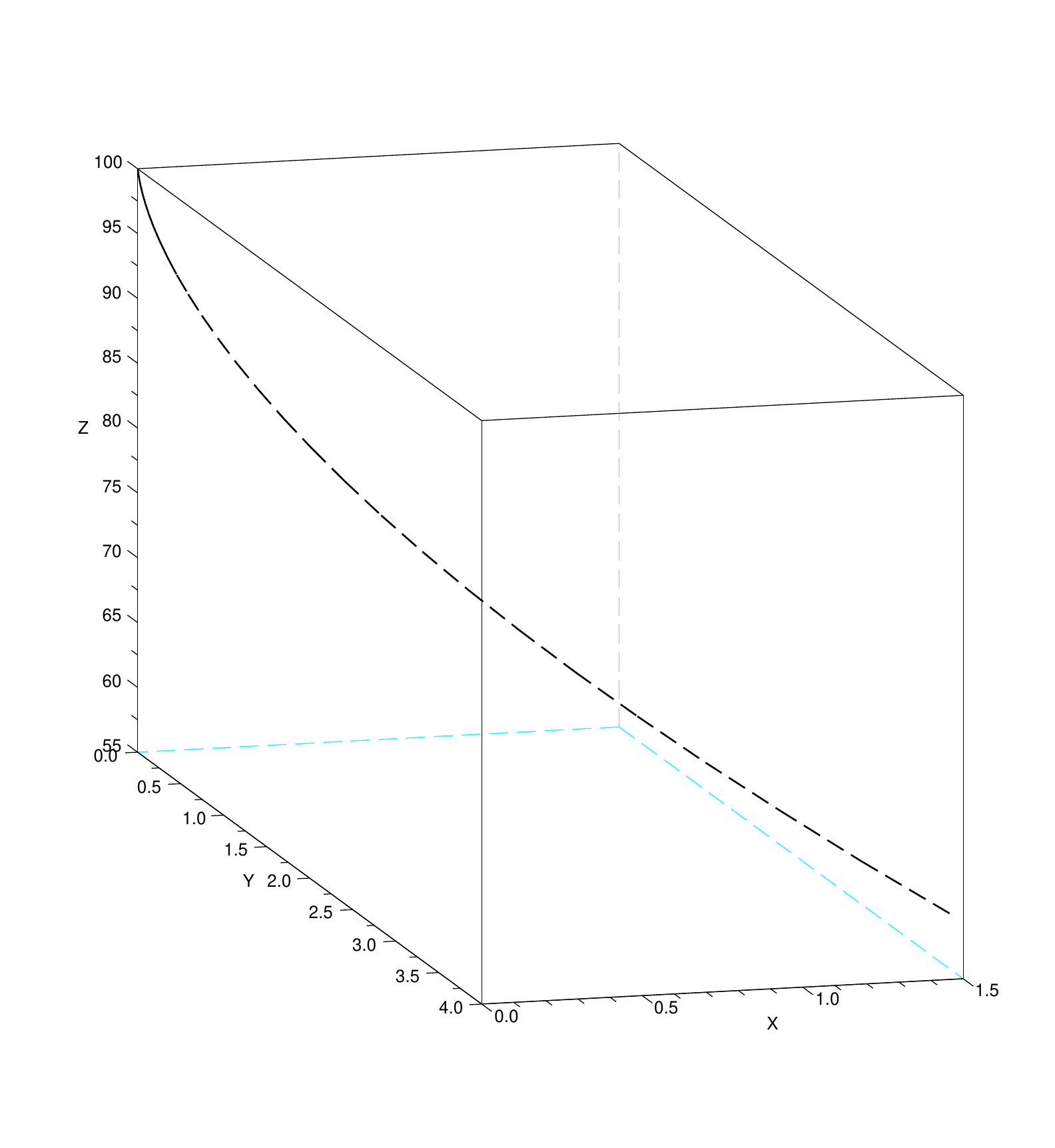}
\caption{Trajectory of a body falling under the influence of the Coriolis force and the air friction 
($\lambda = 45^{\circ}$ N). The altitude $z$ has been multiplied by 0.01, while the $x$ coordinate 
has been amplified by a factor 500; the body is initially at rest.}
\label{fig:bdfall} 
\end{figure}

\begin{figure}[htb]
\centering
\includegraphics[height=6cm]{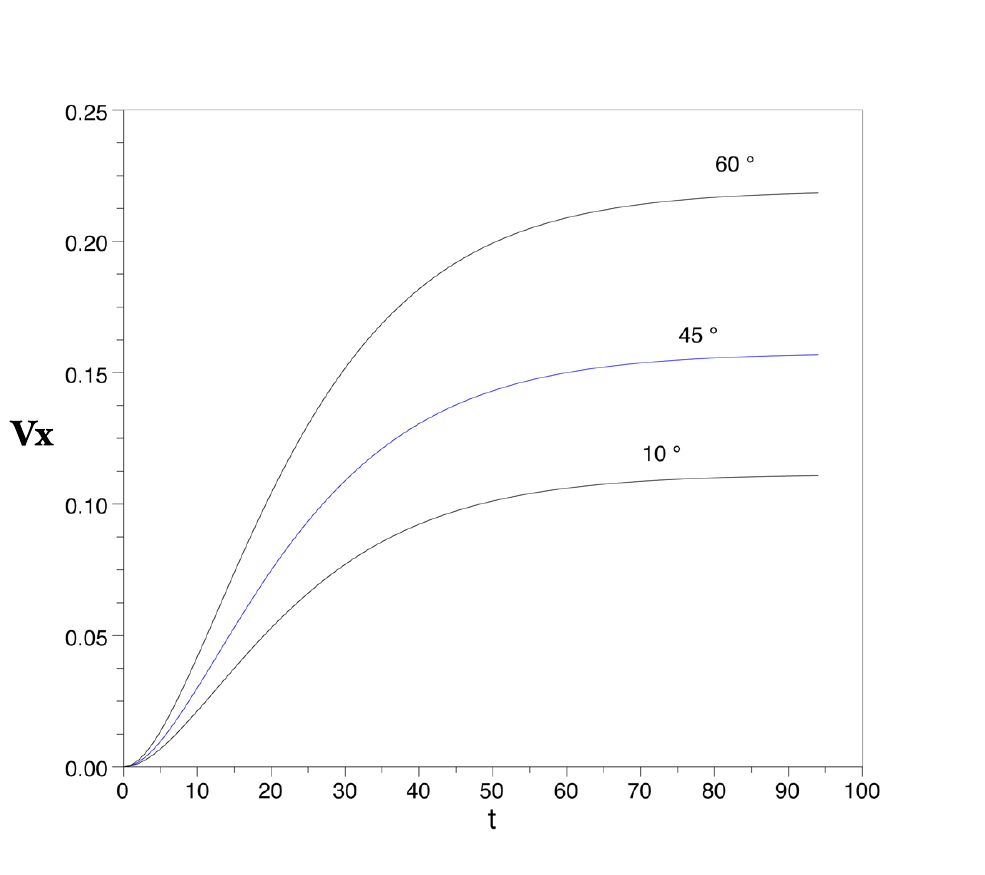}
\caption{Time dependence of the $x$-component of the velocity for different values of the 
latitude.}
\label{fig:veltim} 
\end{figure}

For the above reasons we will treat in detail the solution of the equation of motion of a body falling on the Earth under the action of the 
Coriolis force. If we include the effect of a velocity-dependent force, due, for example, to air friction, we can write this equation 
as \footnote{The term associated to the centrifugal force has not been included. 
We will comment later on the possibility of including this type of contribution in the 
formalism discussed in this paper.}:
\begin{equation}\label{eq:coriol}
\frac{\mathrm{d}}{\mathrm{d}t}\vec{v} \,=\, -\,2\,\vec{\omega} \times \vec{v} \,-\, \eta\,\vec{v} \,+\, \vec{g}\;, 
\qquad \qquad (\vec{\omega} \,=\, \omega\,\vec{n})\;,
\end{equation}
where $\omega$ is the angular velocity of Earth rotation. Also in this case, with an obvious redefinition of the vectors, the formal 
solution is given by eq. (\ref{eq:solcnh}). The explicit form for the velocity vector is: 
\begin{eqnarray}
\label{eq:corvel}
\vec{v} (t) \!\!&=&\!\! \mathrm{e}^{-\eta t}\,\cos (2 \omega t)\,\vec{v}_0 \,+\,\left\{a \,-\, \mathrm{e}^{-\eta t}\,
                          \left[a\,\cos (2 \omega t) \,-\, b\,\sin (2 \omega t)\right]\right\} \vec{g}  \nonumber \\
                      & &\qquad\qquad\qquad\;\;\;+\, (\vec{n} \cdot \vec{\ell}\,)\,\vec{n} \,-\, \vec{n} \times \vec{m}  \;,
\end{eqnarray}
where
\begin{eqnarray}
\vec{\ell}  \!\!&=\!\!& \mathrm{e}^{-\eta t}\,\left[1 \,-\, \cos (2 \omega t)\right] \vec{v}_0 \,+\, 
                                   \left\{\frac{1}{\eta} \,-\, a \,-\, \mathrm{e}^{-\eta t}\,\left[\frac{1}{\eta} \,-\, 
                                   a\,\cos (2 \omega t) \,+\, b\,\sin (2 \omega t)\right] \right\} \vec{g}\;, \nonumber \\
\vec{m}  \!\!&=\!\!& \mathrm{e}^{-\eta t}\,\sin (2 \omega t)\,\vec{v}_0 \,+\, \left\{b \,-\, \mathrm{e}^{-\eta t}\,
                                  \left[a\,\sin (2 \omega t) \,+\, b\,\cos (2 \omega t)\right]\right\} \vec{g}\;, \nonumber \\
                    &&  \qquad\qquad\qquad          a \,=\, \frac{\eta}{\eta^2 + 4 \omega^2}\;, \qquad 
                                    b \,=\, \frac{2 \omega}{\eta^2 + 4 \omega^2}\;, \nonumber 
\end{eqnarray}
while, for the position vector one has 
\begin{eqnarray}
\vec{r} (t)\!\!&=&\!\! \vec{r}_0 \,+\, \left\{a \,-\, \mathrm{e}^{-\eta t}\,\left[a\,\cos (2 \omega t) \,-\, b\,\sin (2 \omega t)\right]
                                \right\} \vec{v}_0  \nonumber \\
                    & &\quad+\, \left\{a t \,-\, a^2 \,+\, b^2 \,+\, \mathrm{e}^{-\eta t}\, \left[(a^2 \,-\, b^2)\,\cos (2 \omega t) 
                           \,-\, 2 a b\, \sin (2 \omega t)\right]\right\}\vec{g} \\
                     & &\quad+\,  \left(\vec{n} \cdot \vec{o} \right) \vec{n} \,-\, 
                           \vec{n} \times \vec{p} \;,
\end{eqnarray}
where
\begin{eqnarray}
\vec{o}  \!\!&=\!\!& \left\{\frac{1}{\eta} \,-\, a \,-\, \mathrm{e}^{-\eta t}\,\left[\frac{1}{\eta} \,-\, 
                                 a\,\cos (2 \omega t) \,+\, b\,\sin (2 \omega t)\right]\right\} \vec{v}_0 \nonumber \\
                    && \,+\,\left\{\left(\frac{1}{\eta} \,-\, a \right) t \,-\, \frac{1}{\eta^2} \,+\, a^2 \,-\, b^2 \right.  \\
                    && \left. \qquad \,+\, \mathrm{e}^{-\eta t}\,\left[\frac{1}{\eta^2} \,-\, (a^2 - b^2)\,\cos (2 \omega t) \,+\, 
                                2 a b\,\sin (2 \omega t) \right] \right\} \vec{g}\;, \nonumber \\
\vec{p}  \!\!&=\!\!& \left\{b \,-\, \mathrm{e}^{-\eta t}\,\left[a\,\sin (2 \omega t) \,+\, b\,\cos (2 \omega t)\right]
                                \right\} \vec{v}_0 \nonumber \\
                     &&    \,+\, \left\{b t \,-\, 2 a b \,+\, \mathrm{e}^{-\eta t}\, \left[(a^2 - b^2)\,\sin (2 \omega t) 
                                \,+\, 2 a b\, \cos (2 \omega t)\right]\right\} \vec{g}\;. \nonumber 
\end{eqnarray}
In Fig. (\ref{fig:bdfall}) we show the trajectory described by a falling body under the influence of Coriolis and friction 
forces, at a latitude of 45$^{\circ}$ N. The time dependence of the component of velocity along the $x$-axis for 
different values of latitude is shown in Fig. (\ref{fig:veltim}).

\begin{figure}[htb]
\centering
\includegraphics[height=7cm]{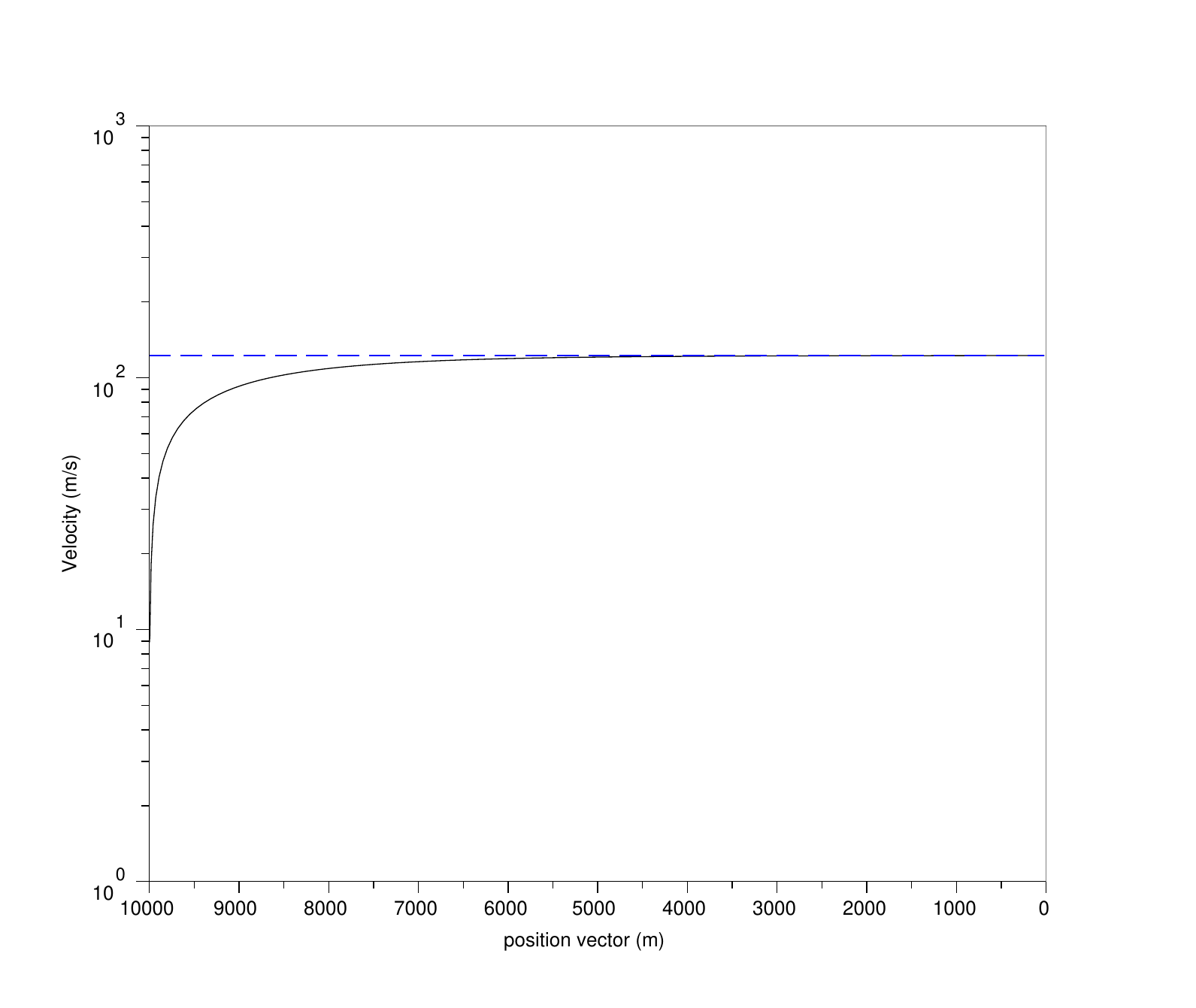}
\caption{The modulus of the velocity vector as a function the modulus of the position vector; 
the dashed line refers to the modulus of the limit velocity vector (same parameters of Fig. (\ref{fig:bdfall})). 
The numerical and analytical solutions yields the same result, thus proving the correctness of the procedure.}
\label{fig:vlim} 
\end{figure}

\begin{figure}[htb]
\centering
\includegraphics[height=7cm]{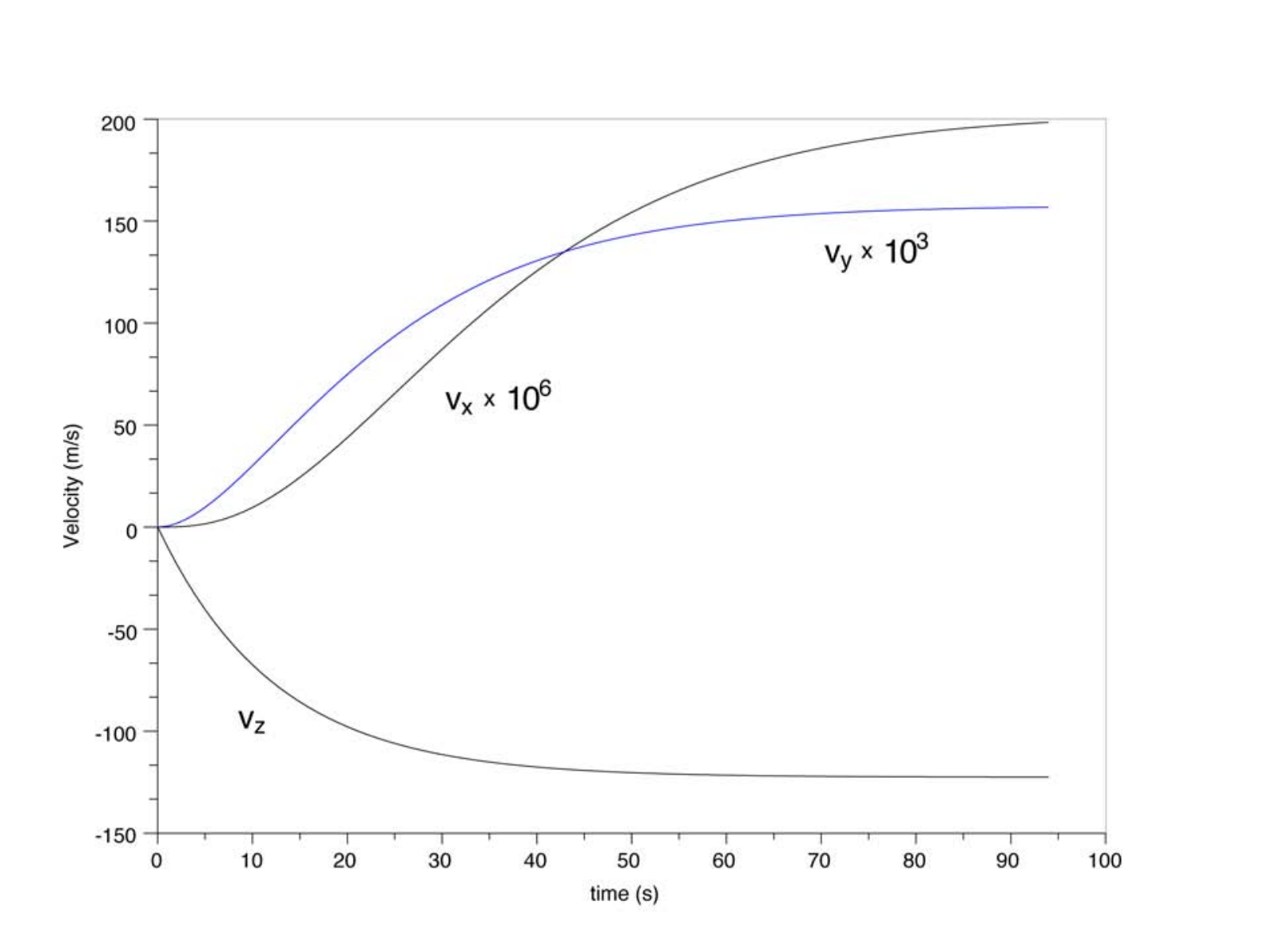}
\caption{The components of the velocity vector as a function of time (same parameters 
of Fig. (\ref{fig:vlim})).}
\label{fig:compv} 
\end{figure}

We expect that after some time the motion will be dominated by the so called limit velocity occurring 
in any problem characterized by a damping term due to friction, and that is reached when the total force 
acting on the moving body is zero. By imposing this condition to the case of eq. (\ref{eq:coriol}), we find 
\begin{equation}
2\,\vec{\omega} \times \vec{v}^{\,\ast}  \,+\, \eta\, \vec{v}^{\,\ast} \,-\, \vec{g} \,=\, 0\;.
\end{equation}
This is a kind of algebraic equation having the velocity vector as the unknown quantity. By applying the 
method developed in this paper, we find for this equation the following formal solution
\begin{equation}
\vec{v}^{\,\ast} \,=\, \frac{1}{\eta + 2 \hat{\omega}}\,\vec{g}\;,
\end{equation}
and taking into account the Laplace transform identity
\begin{equation}
\label{eq:lapla}
\frac{1}{A} \,=\, \int_0^\infty\,\mathrm{d}s\,\mathrm{e}^{-s A}\,,
\end{equation}
valid also if $A$ is an operator, we obtain the following expression for the limit velocity:
\begin{eqnarray}\label{eq:vellim}
\vec{v}^{\,*} \!\!&=&\!\! \int_0^\infty\,\mathrm{d}s\,\mathrm{e}^{- \eta s}\,\mathrm{e}^{-2 s \hat{\omega}}\,\vec{g} 
                            \nonumber \\
               \!\!&=&\!\! \frac{1}{\eta^2 + 4 \omega^2}\,\left[\left(\eta \,+\, 2\,\hat{\omega}\right) \vec{g} \,+\,
                          \frac{4}{\eta}\,\left(\vec{\omega} \cdot \vec{g}\right) \vec{\omega}\right]\;. 
\end{eqnarray}
This expression yields the generalization of the limit velocity occurring in one dimensional motions, but it also 
generalizes the definition of the drift velocity. In Fig. (\ref{fig:vlim}) we have reported the modulus of the velocity 
vector as a function of the modulus of the position vector. It is evident that after a very rapid growth, the velocity 
of the system reaches a limit value consistent with eq. (\ref{eq:vellim}).  In Fig. (\ref{fig:compv}) we have reported 
the components of the velocity vector as function of the time. The effect of the Coriolis force is evident from the 
appearence of the components $v_x$ and $v_y$, which have been significantly amplified to make them visible 
on the same scale for the component $v_z$. 

The results obtained in the present and the previous sections are valid if the vectors defining the VOPs do not 
depend on time and space coordinates. If is not the case, we have a completely different phenomenology, that 
will be discussed in the next sections. 

\section{Field inhomogeneities and the Lorentz-type equations} \label{sec:inho}
Before getting into the specific details relevant to the solution of the equation of Lorentz-type for space and/or 
time dependent fields, it is important to remind  some important issues regarding operational ordering problems 
emerging from the non-commutative nature of the vector product. 

We will illustrate the associated difficulties considering, as an example, the Lorentz equation in absence of electric 
field and with a magnetic field  consisting of two (non parallel) components $\vec{B} = \vec{B}_1 + \vec{B}_2$. 
We may wonder whether the Lorentz equation can be solved in such a way that the solution reflects the above 
decomposition. The evolution operator can be expressed in terms of the VOPs $\hat{\Omega}_{1,2}$  associated 
with the two components of the magnetic field, as $\hat{U} = \exp{[t (\hat{\Omega}_1 + \hat{\Omega}_2)]}$, but, 
as a consequence of the non-commutative character of the vector product, the exponential function does not possess 
the semi-group property $\exp(A + B) = \exp(A) \exp(B) = \exp(B) \exp(A)$ \cite{wilco}.  More sophisticated 
disentanglement procedures of the exponential are in order, as, for example, the Zassenhaus identity \cite{makaso}, 
which provides a formula yielding a disentangled expression for the exponential $\exp(\hat{A} + \hat{B})$ in terms of 
successive commutators of $\hat{A}$ and $\hat{B}$, namely
\begin{equation}\label{eq:zasse}
\mathrm{e}^{\hat{A} + \hat{B}} \,=\, \mathrm{e}^{\hat{A}}\,\mathrm{e}^{\hat{B}}\,\mathrm{e}^{\hat{Z}_1}\,
\mathrm{e}^{\hat{Z}_2} \cdots 
\end{equation}
with
$$
\qquad \hat{Z}_1 \,=\, -\frac{1}{2}\,\left[\hat{A},\,\hat{B}\right]\;, \qquad 
\hat{Z}_2 \,=\, \frac{1}{6}\,\left(2 \left[\hat{B}, \left[\hat{A},\,\hat{B}\right]\right] \,+\, 
\left[\hat{A}, \left[\hat{A},\,\hat{B}\right]\right]\right)\;. \nonumber
$$
The evolution operator containing the sum of two non parallel VOPs cannot be na\"{\i}vely disentangled and,   
accordingly, the associated dynamics cannot be simply expressed as two successive R. r.. The identity 
(\ref{eq:zasse}) can be applied and we will illustrate its usefulness by keeping only the $\hat{Z}_1$  
correction
\begin{equation}
\hat{Z}_1 \,=\, - \frac{t^2}{2}\,\left[\hat{\Omega}_1,\,\hat{\Omega}_2\right] \,=\, 
- t^2 \,\vec{\Omega}_1 \times \vec{\Omega}_2\;, \nonumber
\end{equation}
and, thus, writing the evolution operator as follows\footnote{The symmetric split disentanglement 
$\exp [t (\hat{\Omega}_1 + \hat{\Omega}_2)] \,\simeq\,\exp (\frac{t}{2} \hat{\Omega}_1)\, 
\exp (t\,\hat{\Omega}_2)\, \exp (\frac{t}{2} \hat{\Omega}_1)$ yields an integration more 
accurate than that provided by eq. (\ref{eq:evoapp}) because it is of the order $O(t^3)$. 
The inclusion of further orders in the Zassenhaus expansion may gives a better 
approximation. The symmetric split provides an easier interpretation in geometrical terms, 
since it can be understood as three successive R. r.}
\begin{equation}\label{eq:evoapp}
\mathrm{e}^{t (\hat{\Omega}_1 + \hat{\Omega}_2)} \,\simeq\, \mathrm{e}^{t \hat{\Omega}_1}\,
\mathrm{e}^{t \hat{\Omega}_2}\,\mathrm{e}^{- t^2 (\vec{\Omega}_1 \times \vec{\Omega}_2 )}\;.
\end{equation}
The truncation of the Zassenhaus formula at the first commutator holds for quasi-parallel vectors and/or for 
small times, i.e. when the following inequality is satisfied
\begin{equation}\label{eq:appzas}
 t^2 |\vec{\Omega}_1 \times \vec{\Omega}_2 | \,\ll \, 1\;,
\end{equation}
and, therefore, the approximate solution for the velocity vector can be written as
\begin{equation}
\vec{v} (t) \,\simeq\, \mathrm{e}^{t \hat{\Omega}_1}\,\mathrm{e}^{t \hat{\Omega}_2}\,\vec{v}^{\,\prime} (t)
\end{equation}
where, on account of the condition (\ref{eq:appzas}) and eq. (\ref{eq:svect}), $\vec{v}^{\,\prime}$ can be written as
\begin{equation}\label{eq:vprime}
\vec{v}^{\,\prime} (t) \,\simeq\, \mathrm{e}^{- t^2 (\vec{\Omega}_1 \times \vec{\Omega}_2 )}\,\vec{v}_0 \,\simeq\,
\vec{v}_0 \,-\, \omega_{c,1}\,\omega_{c,2}\,t^2\,\left(\vec{n}_1 \times \vec{n}_2\right) \times \vec{v}_0\;, 
\end{equation}
with
\begin{equation}
\omega_{c,k} \,=\, \frac{e B_k}{m}\;,\qquad \vec{n}_k \,=\, \frac{\vec{B}_k}{B_k} \qquad\qquad (k \,=\, 1,2)\;.
\end{equation}
The successive action of the exponential operators is that of providing two consecutive R. r. of the vector  
$\vec{v}^{\,\prime}$.

For example, if we consider the motion of a charged particle under the action of the terrestrial magnetic field, 
gravity and Coriolis force, we should write the equations of motion as
\begin{equation}
m\,\frac{\mathrm{d}}{\mathrm{d}t}\vec{v} \,=\,- (e\,\vec{B}_T \,+\, 2\,m\,\vec{\omega}) \times \vec{v} 
\,+\, m\,\vec{g}\;,
\end{equation}
where $\vec{B}_T$ denotes the terrestrial magnetic field. According to the previous discussion, the above 
equation can be solved by introducing a kind of equivalent magnetic field vector given by
\begin{equation}\label{eq:bequiv}
\vec{B}^\ast \,=\, \vec{B}_T \,+\, 2\,\frac{m}{e}\,\vec{\omega}\;,
\end{equation}
thus getting for the associated drift velocity the expression
\begin{equation}
\vec{v}_d \,=\, \frac{m}{e}\,\frac{\vec{g} \times \vec{B}^\ast}{B^{\ast 2}}\;,
\end{equation}
that give rise to a current flow orthogonal to the gravity force line and to the direction of the equivalent magnetic force 
(\ref{eq:bequiv}). However, if we are interested to disentangle the magnetic and Coriolis components we can follow 
the just outlined procedure. By assuming that Coriolis and Lorentz force vectors are quasi-parallel, i.e.  
$\vec{B}_T \times \vec{\omega} \simeq 0$,  we obtain a first correction induced by the combined action of the fields 
given by (see eq. (\ref{eq:vprime}))
\begin{equation}
\frac{|\vec{v}^{\,\prime}\,-\,\vec{v}_0|}{v_0} \,\simeq\, \omega_{c,T}\,\omega\,t^2\,\sin \lambda \,\sin \chi\,,
\end{equation}
where $\omega_{c,T}$ is the cyclotron frequency associated to $\vec{B}_T$, $\lambda$ is the angle between the 
vectors $\vec{B}_T$ and $\vec{\omega}$, and $\chi$ the angle between the vector $\vec{B}_T \times \vec{\omega}$ 
and the initial velocity $\vec{v}_0$.

\begin{figure}[htb]
\centering
\includegraphics[height=6cm]{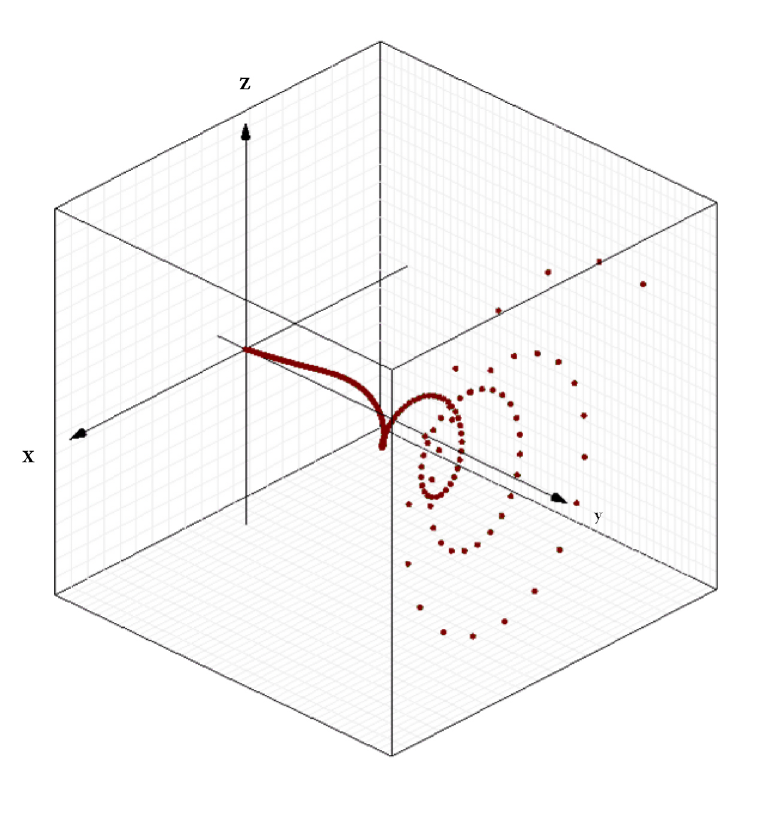}
\caption{Electron motion in a smoothly varying magnetic field lying in the $(x, y)$ plane;
 $\vec{B} =  B_0 (x, -y, 0), \vec{v}_0 = v_0 (0, 1, 0)$.}
\label{fig:nonhom}
\end{figure}

\begin{figure}[htb] 
\centering
\includegraphics[height=5cm]{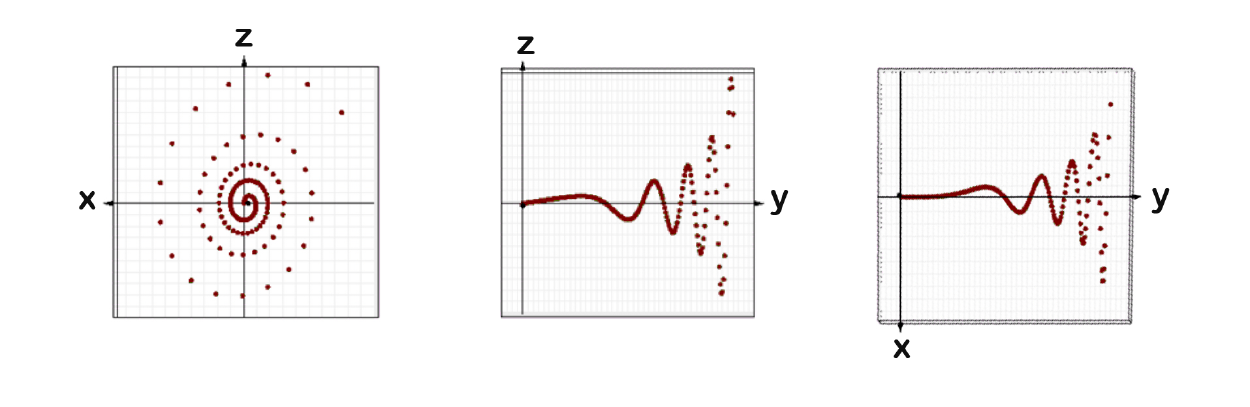}
\caption{Projection of the motion of Fig. (\ref{fig:nonhom}) in:  $(z, x)$ plane (left), 
$(y, z)$ plane (middle) and $(x, y)$ plane (right). The scales are not same (the $y$-coordinates have been 
reduced by a factor 100) to evidentiate the displacement effect.}
\label{fig:project}
\end{figure}

Let us now assume that the magnetic field is not homogeneous, i.e. it exhibits a dependence on the transverse 
coordinates. This modifies the dynamics of the charge undergoing the Lorentz force effect since during its 
motion the particle experiences space regions with different magnetic field intensity and orientation. Furthermore, by 
assuming that we can choose a (small) finite time integration step $\delta$ during which the fields remains constant, 
the velocity and position vectors can be followed, step by step, by means of the following equations (see 
eqs. (\ref{eq:veloc}, \ref{eq:posiz}) with $\vec{Q} = 0$)
\begin{eqnarray}\label{eq:nsteps}
\vec{v}_{k + 1} \!\!&=&\!\! \cos (\Omega_k\,\delta)\,\vec{v}_k \,+\, [\vec{n}_k \cdot \vec{\ell}_k (\delta)]\,
\vec{n}_k \,-\, \vec{n}_k \times \vec{m}_k (\delta)\;, \nonumber \\
\vec{r}_{k + 1} \!\!&=&\!\! \vec{r}_k \,+\,\frac{\sin (\Omega_k\,\delta)}{\Omega_k}\,\vec{v}_k \,+\,
[\vec{n}_k \cdot \vec{o}_k (\delta)]\,\vec{n}_k \,-\, \frac{1}{\Omega_k}\, \vec{n}_k \times 
\vec{\ell}_k (\delta)\;,
\end{eqnarray}
where the index $k$ corresponds to successive integration steps, and
$$
\vec{\ell}_k (\delta) \,=\, \left[1 - \cos (\Omega_k\,\delta)\right]\, \vec{v}_k \qquad 
\vec{o}_k (\delta) \,=\, \left[\delta \,-\, \frac{\sin (\Omega_k\,\delta)}{\Omega_k}\right]\,\vec{v}_k 
$$
$$ 
\vec{B}_k \,=\, \vec{B} (\vec{r}_k)\;, \qquad \vec{\Omega}_k \,=\, \frac{e}{m}\,\vec{B}_k\;.
$$
We report in Figs. (\ref{fig:nonhom}, \ref{fig:project}) an example of motion in a non homogeneous magnetic field, 
where is evident that the main effect of such a coordinate dependence is the appearance of a drift velocity 
contribution \cite{alfv}.

The physical origin of the drift is in the fact that an increase or a reduction of the magnetic field implies 
a corresponding reduction or increase of the Larmor radius, and, therefore, in one period the orbit described by the 
particle is no more closed and the particle, according to the sign of its charge, drifts along the varying field direction. 
A well known, straightforward, calculation allows the evaluation of the drift force under the assumption that the field 
does not vary significantly over a Larmor radius. The magnetic field dependence on the vector position yields 
\begin{equation}
\vec{B} (\vec{r}_0 + \delta \vec{r}) \,=\, \mathrm{e}^{\delta \vec{r} \cdot \vec{\nabla}}\,\vec{B} (\vec{r}_0) 
\,\simeq\, \vec{B}_0 \,+\, (\delta \vec{r} \cdot \vec{\nabla})\vec{B}_0 
\qquad\qquad (\vec{B}_0 = \vec{B} (\vec{r}_0))\,.
\end{equation}
The extra-contribution  to the Lorentz force is
\begin{equation}
\vec{F} \,=\,- e\,< (\delta \vec{r} \cdot \vec{\nabla}) \vec{B}_0 \times \vec{v} >\;,
\end{equation}
where the average is taken over one cyclotron period. In the case $\vec{B}_0\,\perp\,\vec{v}_0$, and assuming 
$\vec{B} \simeq \vec{B}_0$, from eqs. (\ref{eq:veloc}, \ref{eq:posiz}) one obtains 
\begin{equation}
\vec{F} \,\simeq\, -\, \frac{e\,B_0}{2}\,\left[(\vec{n} \times \vec{r}_L) \cdot \vec{\nabla}\right] \,\vec{v}_T\;, 
\qquad\qquad \vec{v}_T \,=\, \vec{n} \times \vec{v}_0\;,
\end{equation}
and the associated drift velocity is evaluated according to eq. (\ref{eq:genvd}). Other types of drift can be included, 
but the procedure remains the same.

\section{Time-dependent fields and Lorentz-type equations}\label{sec:timdep}
The electric and magnetic fields in the Lorentz equation of motion may be time-dependent. The solution of the problem 
in the most general cases presents various difficulties associated to the fact that the field vectors evolve in time and may 
not be parallel to themselves at different times. This situation is reminescent of what occurs in quantum mechanics when 
the hamiltonian is time-dependent and, thus, does not commute with itself at different times. In this case the solution of the 
problem demands for the use of the ordering methods, which can also be exploited for the present problem.

However, let us start with the case in which the fields are time-dependent but their evolution implies only a variation in the 
modulus but not in the direction. Under this hyphotesis time-ordering techniques are not necessary, but the problem deserves 
some comments reported below. We will indeed solve the Lorentz equation of motion for a charge moving in two mutually 
orthogonal electric and magnetic fields, with a sinusoidal time dependence, i.e. we assume\footnote{With this choice 
the Lorentz equation we are going to study are relevant for the motion of a charged particle in the 
field of an electromagnetic wave. We have not used the
wave-like form $\sin(kz - \omega t)$ because if we limit the analysis to the non-relativistic case, 
$kz \ll \omega t$ (see also ref. \cite{rothm}).}
\begin{equation}\label{eq:fields}
\vec{E} \,=\, E_0 \sin (\omega t + \varphi)\,\vec{e}_x\;, \qquad\qquad
\vec{B} \,=\, B_0 \sin (\omega t + \varphi)\,\vec{e}_y\;.
\end{equation}

According to eq. (\ref{eq:fields}) the field vectors remain mutually orthogonal and parallel to themselves at any time 
and, therefore, the solution of the equation is given by eq.  (\ref{eq:solcnh}) with the evolution operator
\begin{equation}
\hat{U} (t) \,=\, \mathrm{e}^{\hat{\Phi} (t)}\,
\end{equation}
where
\begin{equation}
\vec{\Phi} (t) \,=\, \frac{1}{\omega}\,\left[\cos(\omega t + \varphi) \,-\, \cos\varphi \right] \vec{\Omega}_0 
\qquad\qquad\qquad \left(\vec{\Omega}_0 = \frac{e}{m}\,B_0\,\vec{e}_y\right)\;. \nonumber
\end{equation}
Albeit trivial from the mathematical point of view, we report the solution in the simplified case $\vec{E} = 0$
because it presents some aspects useful for next developments ($\vec{n} = \vec{\Omega}_0/\Omega_0$):
\begin{equation}\label{eq:magtdep}
\vec{v} (t) \,=\, \cos (\Phi (t)) \vec{v}_0  \,+\, \left[1 \,-\, \cos (\Phi (t))\right] (\vec{n} \cdot \vec{v}_0) \vec{n} 
\,-\, \sin (\Phi(t))\,\vec{n} \times \vec{v}_0\;.
\end{equation}  
By assuming $ \varphi = \pi/2$, $\vec{n} \cdot \vec{v}_0 = 0$, and using the Jacobi-Anger expansion \cite{andre}, 
one obtains
\begin{eqnarray}
\label{eq:besse}
\cos (x \sin \theta) \!\!&=&\!\!  J_0 (x) \,+\, 2\,\sum_{n = 1}^\infty \, \cos (2 n \theta)\,J_{2 n} (x)\;, \nonumber \\
\sin (x \sin \theta)   \!\!&=&\!\!  2\,\sum_{n = 0}^\infty \,\sin [(2 n + 1)  \theta]\,J_{2 n + 1} (x)\;, 
\end{eqnarray}
we obtain, from eq. (\ref{eq:magtdep}), the following expression for the velocity ($\zeta = \Omega_0/\omega$)
\begin{eqnarray}
\vec{v} (t) \!\!&=&\!\!  J_0 (\zeta)\,\vec{v}_0 \,-\, 2\,J_1 (\zeta)\,\sin (\omega t)\,\vec{n} \times \vec{v}_0 \nonumber \\
&& \quad \,+\, 2\,\sum_{n = 1}^\infty\,\left\{\cos (2 n \omega t)\,J_{2 n} (\zeta)\,\vec{v}_0 \,-\,
\sin [(2 n + 1)  \omega t]\,J_{2 n + 1} (\zeta)\,\vec{n} \times \vec{v}_0\right\}\;,
\end{eqnarray}
which shows the interplay between the cyclotron frequency and the frequency of the oscillating magnetic field. 

The inclusion of the electric field implies some additional computational problems. Again in the case $ \varphi = \pi/2$, 
we write the inhomogeneous term of the solution as follows 
\begin{eqnarray}
\vec{w} (t) \!\!&=&\!\! \mathrm{e}^{\hat{\Phi} (t)}\,\int_0^t\,\mathrm{d}t^{\prime}\,\mathrm{e}^{- \hat{\Phi} (t^{\prime})}\,
\vec{Q} (t^{\prime}) \nonumber \\
           &=&\!\! \mathrm{e}^{- \sin(\omega t)\,\hat{\zeta}}\,\int_0^t\,\mathrm{d}t^{\prime}\,\mathrm{e}^{\sin(\omega t^{\prime})\,
           \hat{\zeta}}\,\cos (\omega t^{\prime})\,\vec{Q}_0  \qquad\qquad \left(\vec{Q}_0 \,=\, \frac{e E_0}{m}\,\vec{e}_x\right)\;.
\end{eqnarray}
This integral can be treated in different ways, and, to have an idea of the mathematical problem one may face with, 
we choose to carry out the integration using the Bessel function expansion of eq. (\ref{eq:besse}). We get:
\begin{equation}
\vec{w} (t)\,=\, \frac{t}{2}\,\sum_{k = - \infty}^\infty\,\sum_{n = - \infty}^\infty\, \exp \left(\imath\,\frac{n + 2k}{2}\,\omega t\right)\, 
F_n (t)\, \vec{J}_{k,n}
\end{equation}
where
\begin{equation}
F_n (t) \,=\, \mathrm{e}^{\imath \omega t/2}\,{\rm sinc} \left(\frac{n + 1}{2}\,\omega t\right) \,+\, 
\mathrm{e}^{- \imath \omega t/2}\,{\rm sinc} \left(\frac{n - 1}{2}\,\omega t\right)\;.\nonumber
\end{equation}
and 
\begin{equation}
\vec{J}_{k,n} \,=\, J_k(\imath \hat{\zeta})\,J_n (- \imath \hat{\zeta})\,\vec{Q}_0\;.
\end{equation}
The last vector can be specified either in terms of the series expansion for the Bessel functions\footnote{
$J_n (z) \,=\, \sum_{k = 0}^\infty \, \frac{(-)^k}{k! (n + k)!}\,(z/2)^{n + 2 k}\,$ 
.} 
or by the use of the integral representation \cite{andre}
\begin{equation}
J_n (x) \,=\, \frac{1}{2 \pi}\,\int_{- \pi}^{\pi}\,\mathrm{d}\theta\, \mathrm{e}^{\imath (x \sin \theta - n \theta)}
\end{equation}
that allows to write 
\begin{equation}
\vec{J}_{k,n} \,=\, \frac{1}{4 \pi^2}\,\int_{- \pi}^{\pi}\,\mathrm{d}\theta\,\int_{- \pi}^{\pi}\,\mathrm{d}\chi\,
\mathrm{e}^{- \imath (k \theta + n \chi)}\,\mathrm{e}^{- (\sin \theta - \sin \chi)\,\hat{\zeta}}\,\vec{Q}_0\;.
\end{equation}

We have mentioned this specific problem to give a very first idea of the problems associated with the 
solution of equations of the type (\ref{eq:lorentz}), where the torque and inhomogeneous vector are explicitly 
dependent on the integration variation, but also because it has interesting implications for the understanding 
of the role played by the Poynting vector in the dynamics of charged particles moving under the combined 
action of mutually time dependent orthogonal fields. The method of solution we have proposed,  having an 
intrinsic vector nature, can be ideally suited to treat the question addressed in \cite{rothm} and clarify the link 
between the Poynting vector and the drift velocity term given in eq. (\ref{eq:vdrift}). This aspect of the problem 
will be treated elsewhere.

\subsection{Operational Methods and time ordering techniques}
Before considering problems requiring time ordered products, we will present a method of solution involving an 
elaboration of the Heaviside operational method \cite{heavi}. To better appreciate the usefulness of the procedure 
we consider the case in which the particle is initially at rest and only the electric field is a function of time while 
the magnetic field is static. We can write the formal solution of eq. (\ref{eq:lorentz}) in the following way
\begin{equation}\label{eq:voper}
\vec{v} (t)\,=\, \left(\frac{\mathrm{d}}{\mathrm{d}t} \,+\, \hat{\Omega}\right)^{- 1} \, \vec{Q} (t)\;,
\end{equation}
and the use of the Laplace transform methods leads to
\begin{equation}
\vec{v} (t)\,=\, \int_0^\infty\,\mathrm{d}s\, \exp\left\{- s \left(\frac{\mathrm{d}}{\mathrm{d}t} \,+\, 
\hat{\Omega}\right)\right\}\, \vec{Q} (t)\;.
\end{equation}
In this integral the exponential can be straightforwardly disentangled because the VOP and derivative operators 
appearing in its argument commute between them. Therefore, from eq. (\ref{eq:veloc}), under the further assumption 
that the electric field differs from zero only for $t > 0$, one has ($\vec{n} = \vec{\Omega}/\Omega$)
\begin{equation}
\vec{v} (t)\,=\, \int_0^t\,\mathrm{d}t^{\prime}\,\mathrm{e}^{(t^{\prime} - t) \hat{\Omega}}\, \vec{Q} (t^{\prime}) \,=\,
\vec{c} \,+\, (\vec{n} \cdot \vec{f}\,)\,\vec{n} \,+\, \vec{n} \times \vec{s}\;, 
\end{equation}
with
\begin{eqnarray}
\vec{c} \!\!&=&\!\!  \int_0^t\,\mathrm{d}t^{\prime}\,\cos \left[\Omega (t^{\prime} - t)\right]\, \vec{Q} (t^{\prime})\;,
\qquad \qquad \vec{s} \,=\, \int_0^t\,\mathrm{d}t^{\prime}\,\sin \left[\Omega (t^{\prime} - t)\right]\, 
\vec{Q} (t^{\prime})\;, \nonumber \\
&& \qquad \vec{f} \,=\,  \int_0^t\,\mathrm{d}t^{\prime}\,\left\{1 \,-\, \cos \left[\Omega (t^{\prime} - t)\right]\right\}\, 
\vec{Q} (t^{\prime})\;.
\end{eqnarray}

The case in which the vector $\vec{\Omega}$ varies with time, not only in modulus but also in direction, in such 
a way that 
\begin{equation}
\label{eq:omvect}
\vec{\Omega}(t_1)\,\times\, \vec{\Omega}(t_2)\, \neq 0\;, 
\end{equation}
implies that the VOPs associated at different times do not commute. The solution of our problem cannot be obtained
using a straightforward integration of the time dependent part, i.e. it will not be sufficient replace in the Rodrigues rotation  
$t\,\hat{\Omega}$ with $\int_0^t\,\mathrm{d}\tau\, \hat{\Omega}(\tau)$. 

From the geometrical point of view the condition (\ref{eq:omvect}) states that the torque vector is no more parallel to itself 
at different times, or that the corresponding matrix equation is expressed in terms of an explicitly time-dependent matrix, 
not commuting with itself at different times. The situation is clearly reminiscent of what is occurring in quantum mechanics 
where the solution of Schr\"{o}dinger problems requires a time ordered expansion of the evolution operator, like the Dyson 
\cite{dyson} or Magnus \cite{magnu} expansion. We will treat the problem by exploiting the theory of 
\emph{path ordered exponential} \cite{datgato}.

From the mathematical point of view the path ordered exponential function is defined in non commutative fields and is equivalent 
to the exponential function in a commutative field. We define, therefore, the following ordered exponential with respect to the 
ordering parameter $t$:
\begin{equation}
\mathrm{oe}[\hat{T}](t) \,=\, \left(\exp\left\{\int_0^t\,\mathrm{d}t^{\prime}\,\hat{T}(t^{\prime})\right\}\right)_+
\end{equation}
where the symbol $( )_+$ denotes the Dyson time ordering operator for the element $\hat{T}(t)$ of an algebra with a non 
commutative product $\circ$. The ordered exponential can be defined in many different ways. Here we use the differential equation
\begin{equation}
\label{eq:oexp}
\frac{\mathrm{d}}{\mathrm{d}t} \mathrm{oe}[\hat{T}](t) \,=\,\hat{T}(t)\,\circ\,\mathrm{oe}[\hat{T}](t)\;,
\qquad\qquad (\mathrm{oe}[\hat{T}](0)\,=\,\hat{1})\;.
\end{equation}
The solution of eq. (\ref{eq:oexp}) can be written in terms of the following series
\begin{equation}
\mathrm{oe}[\hat{T}](t) \,=\, \hat{1} \,+\, \sum_{n = 1}^\infty\, \mathrm{oe}_n[\hat{T}](t) 
\end{equation}
with
\begin{equation}
\mathrm{oe}_n[\hat{T}](t) \,=\, \frac{1}{n!}\, \int_0^t\, \mathrm{d}t^{\prime}\, 
\hat{T}(t^{\prime}) \,\circ\, \mathrm{oe}_{n - 1}[\hat{T}](t^{\prime})\;,
\qquad \mathrm{oe}_0[\hat{T}](t) \,=\, \hat{1}\;. \nonumber
\end{equation}
The above solution is clearly recognized as an ordinary Dyson expansion \cite{dyson}, which is not fully satisfactory 
because it is a perturbative series which does not ensure properties, e.g. the conservation of the norm of a vector, holding 
also for time-dependent vectors. Different  expansions, preserving at any order the norm, can be employed as, for 
example, the already quoted Magnus expansion (see also ref. \cite{datgato}), which can be written as follows
\begin{equation}
\mathrm{oe}[\hat{T}](t) \,=\,\exp \left\{\int_0^t\,\mathrm{d}t^{\prime}\,\hat{T}(t^{\prime}) \,+\, 
\frac{1}{2}\, \int_0^t\,\mathrm{d}t_1\,\int_0^{t_1}\,\mathrm{d}t_2\,\left[ \hat{T} (t_1),\,\hat{T} (t_2) \right] \,+\, 
\cdots \right\}\;,
\end{equation}
where the dots refer to higher order commutators, not reported here. Let us note that in the case of Lorentz equation 
the non commutative product is the vector product ($\circ \equiv \times$), and, therefore
\begin{equation}\label{eq:commut}
\left[ \hat{T} (t_1),\,\hat{T} (t_2) \right] \,=\, 
 2\,\vec{T} (t_1) \,\circ\, \vec{T} (t_2)\,.
\end{equation}
According to this result, the solution of the Lorentz equation (\ref{eq:lorentz}) with $\vec{\Omega}$ and $\vec{Q}$ 
depending on time is given by eq. () with the evolution operator 
\begin{equation}
\hat{U} (t) \,=\, \mathrm{e}^{- \hat{\Gamma} (t) + \hat{\Delta} (t)}\;,
\end{equation}
where the following notation has been introduced
\begin{equation}
\hat{\Gamma} (t) \,=\, \int_0^t\,\mathrm{d}\tau\,\hat{\Omega}(\tau) \;, \qquad\qquad
\hat{\Delta} (t) \,=\, \frac{1}{2}\, \int_0^t\,\mathrm{d}t_1\,\int_0^{t_1}\,\mathrm{d}t_2\,
\left[ \hat{\Omega} (t_1),\,\hat{\Omega} (t_2) \right] \,+\, \cdots
\end{equation}

As for the correction $\hat{\Delta} (t)$, if we assume that the modulus of the vector $\vec{\Omega}$ remains constant, 
one has
\begin{equation}
\left[ \hat{\Omega} (t_1),\,\hat{\Omega} (t_2) \right] \,=\,  2\,\Omega^2\,\sin \theta_{12}\,\vec{u}\;, 
\end{equation}
where $\theta_{12}$ is the angle formed by the two vectors and $\vec{u}$ is the versor pointing in the direction 
orthogonal to the plane defined by $\vec{\Omega} (t_1)$ and  $\vec{\Omega} (t_2)$. For sufficiently small time 
differences and for adiabatic changes, we expect that $\theta_{12} = \omega (t_2 - t_1) \ll 1$, and therefore
\begin{equation}\label{eq:deltac}
\hat{\Delta} (t) \,\simeq\, \frac{1}{6}\, \Omega^2\,\omega\,t^3\,\vec{u}\;.
\end{equation}
Higher orders corrections can also be included but calculations becomes more and more cumbersome. 

The previous discussion has been developed on purely mathematical grounds. As an example of application, we can 
consider the motion of a particle under the influence of a magnetic field with a slowly varying component along 
the $z$-axis and a constant $y$-component, i.e.  the magnetic vector changes its direction and it is not 
parallel to itself at any time. The lowest order corrections in the Magnus expansion allow the inclusion of the effect of 
the slow field evolution. If we assume that the variation is adiabatic over one gyration period, the correction 
can be evaluated by means of eq. (\ref{eq:deltac}), with a frequency $\omega$, assumed to be constant during this 
time, given by
\begin{equation}
\omega \,=\, \frac{B_y \dot{B}_z}{B^2}\;.
\end{equation}

\section{Second order Lorentz equation}\label{sec:2ndord}
Before entering the specific topic of this section we will discuss the application of the operational method developed in 
the previous sections to the solution of the following evolution equation
\begin{equation}
\frac{\mathrm{d}}{\mathrm{d}t} \vec{S} \,=\, \vec{T} \times \vec{S} \,+\, \lambda\,\vec{T} \times 
\left(\vec{T} \times \vec{S}\right)\;, \qquad\qquad\bar{S}|_{t = 0}\,=\,\vec{S}_0\;,
\end{equation}
which is an evolution-type vector equation, with a further contribution associated with a double vector product of the 
torque vector. The associated evolution operator writes
\begin{equation}
\hat{U} (t) \,=\, \mathrm{e}^{t\,\hat{T} \,+\, \lambda\,t\,\hat{T}^2}
\end{equation}
which involves linear and quadratic VOPs. It resembles the generating function of two variable Hermite polynomials 
$H_n(a,b)$ \cite{appel} and can be therefore expanded in series according to the identity
\begin{equation}
\mathrm{e}^{a\,\xi \,+\, b\,\xi^2} \,=\, \sum_{n = 0}^\infty \,\frac{\xi^n}{n!}\,H_n (a,b)\, \qquad\qquad 
H_n (a,b) \,=\, n!\,\sum_{k = 0}^{[n/2]}\,\frac{1}{(n - 2k)!\,k!}\,a^{n - 2k}\,b^k\,.
\end{equation}
The use of the operator $\hat{T}$  as expansion parameter yields the following series for the evolution operator
\begin{equation}
\hat{U} (t) \,=\, \sum_{n = 0}^\infty \,\frac{1}{n!}\,H_n (t, \lambda t)\,\hat{T}^n\;.
\end{equation}
Even though slightly more complicated than an ordinary exponential expansion, we can again take advantage from the 
cyclic properties of the vector product to reduce it to a kind of Rodrigues rotation, by defining the following cos- and 
sin-like functions
\begin{equation}\label{eq:triglike}
{\rm Ch} (t) \,=\, \sum_{n = 0}^\infty \,\frac{(-)^n}{(2 n)!}\, T^{2 n}\, H_{2 n} (t, \lambda t)\;, 
\quad\quad {\rm Sh} (t) \,=\, \sum_{n = 0}^\infty \,\frac{(-)^n}{(2 n + 1)!}\, T^{2 n + 1}\, H_{2 n + 1} (t, \lambda t)\,.
\end{equation}
The form of the solution is therefore exactly that given in eq. (\ref{eq:svect}) with the replacements $\cos \to$ Ch, $\sin \to$ 
Sh. In the case in which the evolution operator
\begin{equation}
\hat{U} (t) \,=\, \exp\left\{\sum_{k = 1}^p\,\lambda_k\,t\,\hat{T}^k\right\}
\end{equation}
we can exploit the same procedure involving higher orders Hermite polynomials, as we will discuss in the concluding section.

Let us now discuss the inclusion of radiation correction effects, which are usually incorporated in the classical Lorentz 
equation by means of a second order time derivative term, according to the following expression \cite{jacks}
\begin{equation}\label{eq:radcorr}
\left(-\,\tau\,\frac{\mathrm{d}^2}{\mathrm{d} t^2} \,+\, \frac{\mathrm{d}}{\mathrm{d} t} \,+\, \hat{\Omega}\right)\,\vec{v} 
\,=\, \vec{Q} (t)
\end{equation}
where
\begin{equation}
\vec{v} (0) \,=\, \vec{v}_0 \qquad \frac{\mathrm{d}}{\mathrm{d} t}\,\vec{v}|_{t = 0} \,=\, \vec{a}_0\;, \qquad
\tau  \,=\, \frac{2}{3}\,\frac{r_0}{c}\;, \qquad r_0  \,=\, \frac{e^2}{m c^2}\;. \nonumber
\end{equation}
In this case equation we have the contribution of an extra term that acts as a an anti-damping giving rise to the so called 
runaway solutions. The physical content of this equation is well known and will not be commented here. We will limit our 
analysis to its mathematical aspects, which have some elements of interest, since this equation represents a 
non-homogeneous second order differential vector equation and we can use an extension of the previously outlined method 
to write the relevant solution. For future convenience we factorize the operator acting on the velocity vector as follows
\begin{equation}
\left(\frac{\mathrm{d}}{\mathrm{d}t} \,-\, \hat{A}_+\right)\,\left(\frac{\mathrm{d}}{\mathrm{d}t} \,-\, \hat{A}_- \right)\,\vec{v} 
\,=\, - \frac{1}{\tau}\, \vec{Q}
\end{equation}
where
\begin{equation}
\hat{A}_\pm \,=\, \frac{1}{2 \tau}\,(1 \,\pm\, \hat{\alpha})\;, \qquad\qquad 
\left(\hat{\alpha} \,=\, \sqrt{1 \,+\, 4 \tau \hat{\Omega}}\,\right)\;. \nonumber
\end{equation}
In the hypothesis that $\vec{v}_0 = \vec{a}_0 = 0$,  we can write the formal solution of eq. (\ref{eq:radcorr}) using a 
generalization of eq. (\ref{eq:voper}), namely
\begin{eqnarray}
\vec{v} (t) \!\!&=&\!\! - \frac{1}{\tau}\,\left[\left(\frac{\mathrm{d}}{\mathrm{d}t} \,-\, \hat{A}_+\right)\,
                                   \left(\frac{\mathrm{d}}{\mathrm{d}t} \,-\, \hat{A}_-\right)\right]^{- 1} \, \vec{Q} (t)\; \nonumber \\
                     &=&\!\! - \frac{1}{\tau (\hat{A}_+ - \hat{A}_-)}\,\left\{\left(\frac{\mathrm{d}}{\mathrm{d}t} \,-\, 
                                  \hat{A}_+\right)^{\!\!- 1} - \left(\frac{\mathrm{d}}{\mathrm{d}t} \,-\, \hat{A}_-\right)^{\!\!- 1}\right\}\,\vec{Q} (t)\,,
\end{eqnarray}
from which, by using the Laplace transform identity (\ref{eq:lapla}) and assuming that $\vec{Q} = 0$ for $t < 0$,  we obtain
\begin{equation}
\vec{v} (t) \,=\, - \frac{1}{\hat{\alpha}}\,\left[\mathrm{e}^{t \,\hat{A}_+}\,\int_0^t\,\mathrm{d}\xi\, 
\mathrm{e}^{- \xi\,\hat{A}_+}\,\vec{Q} (\xi) \,-\, \mathrm{e}^{t \,\hat{A}_-}\,\int_0^t\,\mathrm{d}\xi\,
\mathrm{e}^{- \xi \,\hat{A}_-}\,\vec{Q} (\xi)\right]\;.
\end{equation}
We note that 
\begin{equation}
\frac{1}{\hat{\alpha}} \,=\, \frac{1}{\sqrt{\pi}}\,
\int_0^\infty\, \mathrm{d}s\, \frac{\mathrm{e}^{-s\,(1 \,+\, 4\,\tau\,\hat{\Omega})}}{\sqrt{s}}\;,
\end{equation}
and, using the Newton series expansion for the operator $\hat{A}_\pm$, we get for the exponential operators:
\begin{equation}
\label{eq:expop}
\mathrm{e}^{t \hat{A}_\pm} \,=\, \exp \left\{\frac{t}{2 \tau}\,\left[1 \,\pm\, \sum_{k = 0}^\infty \,\binom{1/2}{k}\, 
(4\,\tau\,\hat{\Omega})^k \right]\right\}\;,
\end{equation}
which can be written in terms of a Rodrigues rotation involving the previously quoted sin- and cos-like functions expressed 
in terms of the higher orders Hermite polynomials (see eq. (\ref{eq:triglike})).

According to the previously outlined steps we know how to handle the formal expression given in eq. (\ref{eq:expop}) to get 
an explicit solution for our problem. Let us now consider only the homogeneous part of eq. (\ref{eq:radcorr}), whose formal 
solution reads
\begin{equation}
\vec{v} (t) \,=\, \mathrm{e}^{t \,\hat{A}_+}\,\vec{c}_1 \,+\, \mathrm{e}^{t \,\hat{A}_-}\,\vec{c}_2\;,
\end{equation}
where $\vec{c}_{1,2}$ are constant vectors linked to the initial vectors by the relations
\begin{eqnarray}
(\hat{A}_+ \,-\, \hat{A}_-)\,\vec{c}_1 \!\!&=&\!\! - \hat{A}_-\,\vec{v}_0 \,+\, \vec{a}_0 \nonumber \\
(\hat{A}_+ \,-\, \hat{A}_-)\,\vec{c}_2 \!\!&=&\!\! \;\;\hat{A}_+\,\vec{v}_0 \,-\, \vec{a}_0\;.
\end{eqnarray}
The action of the exponential operators on the constant vectors can be defined according to the previous prescriptions.

Since eq. (\ref{eq:radcorr}) is a second order equation, we can cast it in the form of a matrix equation as follows
\begin{equation}\label{{eq:matrix}}
\frac{\mathrm{d}}{\mathrm{d}t}\,\underline{Z} \,=\, \hat{M}\, \underline{Z} \,+\, \underline{K}\;,
\end{equation}
where
\begin{equation}\label{eq:matrix}
\underline{Z} \,=\,\left(
\begin{array}{c}
      \vec{v}    \\
      \vec{a}  
\end{array}
\right)\;, \qquad 
\hat{M} \,=\,\frac{1}{\tau}\, \left(
\begin{array}{cc}
                     0 & \tau   \\
\hat{\Omega} & 1
\end{array}
\right)\;, \qquad
\underline{K} \,=\, - \frac{1}{\tau}\,\left(
\begin{array}{c}
      0 \\
     \vec{Q}   
\end{array}
\right)\;.
\end{equation}
The evolution operator $\mathrm{e}^{t\,\hat{M}}$ associated to this equation is given by the exponential of a $2 \times 2$ matrix. 
Since the matrix $\hat{M}$ does not depend on time, standard means, e.g. the Cayley-Hamilton theorem \cite{dator}, can be 
used to cast it in the form reported below ($\sigma = t/2 \tau$)
\begin{equation}
\hat{U} (t) \,=\, \mathrm{e}^{t\,\hat{M}}\,=\,
\mathrm{e}^{\sigma}\,\left(
\begin{array}{cc}
   -\frac{1}{\hat{\alpha}}\,\sinh (\sigma\,\hat{\alpha}) \,+\, \cosh (\sigma\,\hat{\alpha})  &   
     2\,\frac{\tau}{\hat{\alpha}}\, \sinh (\sigma\,\hat{\alpha})  \\
    2\,\frac{\hat{\Omega}}{\hat{\alpha}}\,\sinh (\sigma\,\hat{\alpha})  &  
    \frac{1}{\hat{\alpha}}\,\sinh (\sigma\,\hat{\alpha}) \,+\, \cosh (\sigma\,\hat{\alpha})  
\end{array}
\right)\;, 
\end{equation}
and, thus ($\underline{Z}_0 = \underline{Z}|_{t = 0}$)
\begin{equation}
\underline{Z} \,=\, \hat{U} (t)\,\underline{Z}_0 \,+\, \int_0^t\,\mathrm{d}t^\prime \,\hat{U}^{-1} (t^\prime)\, 
\underline{K}\;,
\end{equation}
that is the same solution as before, written in matrix notation.  

The structure of the previous formalism may rises some confusion. As an example, let us consider the definition 
of the norm of  $\underline{Z}$:
\begin{equation}
| \underline{Z} |^2 \,=\, \left(\vec{v}\;\;\vec{a}\right) \,\cdot \left(
\begin{array}{c}
      \vec{v} \\
      \vec{a}   
\end{array}
\right)
\end{equation}
where the dot represents an ordinary scalar product. It is evident that, according to such a definition, the norm is not 
preserved. This is due to the rotation matrix, which is not norm-preserving, and to the runaway mechanism, associated 
with the anti-damping term e$^\sigma$. Such a term is, from the mathematical point of view, not particularly significant 
and can always be eliminated by means of a Liouville transformation \cite{andre}.

Let us consider again the following second order vector equation  
\begin{equation}\label{eq:2ndvect}
\frac{\mathrm{d}^2}{\mathrm{d} t^2}\,\vec{s} \,+\, \hat{A}\,\frac{\mathrm{d}}{\mathrm{d} t}\,\vec{s} \,+\,\hat{B}\,\vec{s} \,=\, 0
\end{equation}
where $\hat{A}$ and $\vec{B}$ are VOPs. An equation of this type is met in physics in the study of Coriolis problems, including 
the effects of the centrifugal forces, and presents an extra difficulty associated with the presence of the vector operator in the 
damping term. Eq. (\ref{eq:2ndvect}) can be reduced to a Liouville standard form by setting
\begin{equation}\label{eq:liouvil}
\vec{s} (t) \,=\, \exp \left(- \frac{t}{2}\,\hat{A}\right)\,\vec{u} (t)\;
\end{equation} 
and, if $[\vec{A}, \vec{B}] = 0$, we can write the equation for the vector $\vec{u}$ as follows
\begin{equation}\label{eq:eqlio}
\frac{\mathrm{d}^2}{\mathrm{d} t^2}\,\vec{u} \,+\, \left(\hat{B} \,-\, \frac{1}{4}\,\hat{A}^2\right)\,\vec{u} \,=\, 0\;.
\end{equation}
The transformation provided by eq. (\ref{eq:liouvil}) is geometrically interpreted as a R. r. of the vector $\vec{u}$, induced by 
the torque vector associated with the VOP $\hat{A}$. The solution of eq. (\ref{eq:eqlio}) can be obtained using the procedure 
illustrated before. 

The previous results (see also the first paper in Ref. \cite{dattoli}) can be exploited to develop a numerical code for the motion of 
bodies under the influence of gravity, Coriolis and centrifugal forces.

\section{Relativistic effects}\label{sec:releff}
\begin{figure}[htb]
\centering
\includegraphics[height=14cm]{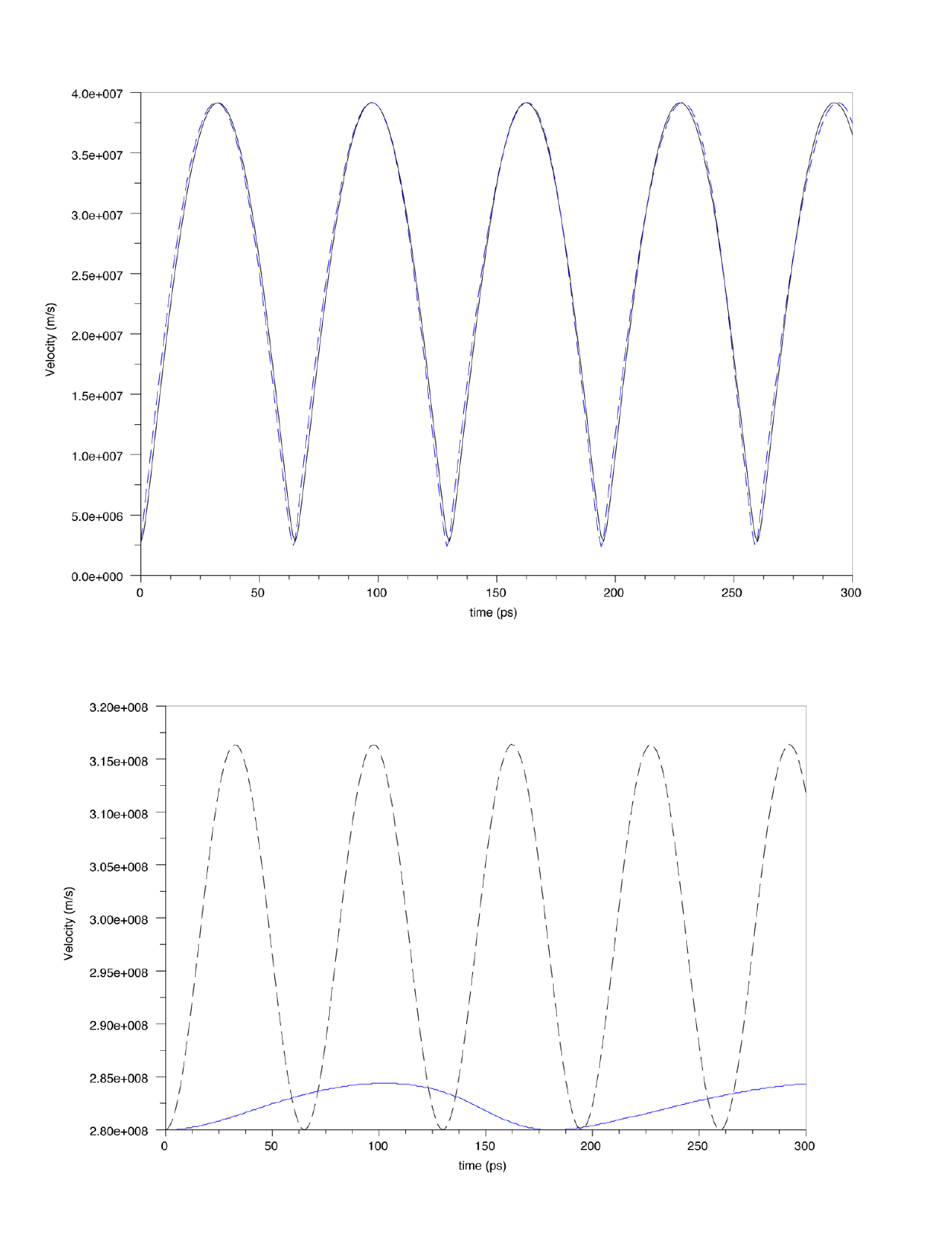}
\caption{ Modulus of the velocity as function of time: comparison between the non relativistic (left) and 
the relativistic (right) case for $\vec{v}_0 = v_0 (1, 0, 0)$; $\vec{B} = B_0 (0, 0, 1)$; $\vec{E} = E_0 (0, -1, 0)$.}
\label{fig:relat} 
\end{figure}

In the previous sections we have discussed the motion of charged particles in electric and magnetic field without considering 
any relativistic correction. This can be easily accounted for by rewriting the Lorentz equation in the form
\begin{equation}\label{eq:rellor}
m_0\,\frac{\mathrm{d}}{\mathrm{d} t}\,\left(\gamma\,\vec{v}\right) \,=\, e\,\left(- \vec{B} \times \vec{v} \,+\,\vec{E}\right)\;
\end{equation}
where $m_0$ is the electron rest mass and $\gamma$ is the relativistic factor. The integration of this equation appears problematic, 
even for constant and homogeneous fields, since $\vec{E}$ induces a non conservation of the modulus of the velocity. The relativistic 
factor $\gamma$  is no more constant and a further equation, specifying its time dependence, should be coupled to eq. (\ref{eq:rellor}).
By introducing the vector $\vec{\Lambda} = \gamma \vec{v}$, eq. (\ref{eq:rellor}) can be rewritten as (with the usual means for the 
vectors $\vec{\Omega}$ and $\vec{Q}$)
\begin{equation}
\frac{\mathrm{d}}{\mathrm{d} t}\,\vec{\Lambda} \,=\, - \frac{1}{\gamma}\, \vec{\Omega} \times \vec{\Lambda} \,+\, \vec{Q}\;,
\end{equation}
By multiplying both sides of this equation by the velocity, we obtain
\begin{equation}
\vec{\Lambda}\,\cdot\,\frac{\mathrm{d}}{\mathrm{d} t}\,\vec{\Lambda} \,=\, \vec{\Lambda}\,\cdot\,\vec{Q}\;.
\end{equation}
and, thus ($\vec{\Lambda}_0 = \vec{\Lambda} (t = 0)$)
\begin{equation}
\Lambda^2 (t) \,-\, \Lambda_0^2 \,=\, 2\,\int_0^t\,\mathrm{d}t^\prime\, \vec{Q}\,\cdot\,\vec{\Lambda} (t^\prime)\;,
\end{equation}
that provides the relativistic kinetic energy variation due to the interaction of the charge with the electric field. 
As already stressed, the presence of the relativistic factor prevents us from the possibility of finding an analytical solution using the tools 
developed so far. However, a fairly straightforward integration scheme can be used adopting the same iterative procedure described in 
sec. \ref{sec:inho} (see eqs. (\ref{eq:nsteps})). In this case the solution is obtained at each time step in terms of eqs. (\ref{eq:veloc}), in 
which $\Omega$ is replaced by $\Omega/\gamma_{n - 1}$. An example of solution of the relativistic problem is shown in Fig. (\ref{fig:relat}) 
where we have plotted the modulus of the velocity as function of time using the solution given in eqs. (\ref{eq:veloc}) and including the 
relativistic correction.
It is evident that for low initial velocities the two solutions are undistinguishable, while the differences becomes significant with increasing 
values of $v_0$. The oscillations in the velocity are due to the combined effect of electric and magnetic fields, but in the relativistic case 
the oscillation period becomes larger, in agreement with the fact that the cyclotron frequency decreases with increasing  $\gamma$.

We do not include further examples because the goal of this section was just showing that the analytical solutions could straightforwardly 
be implemented into a numerical scheme yielding the relativistic treatment. These effects will be thoroughly discussed in a forthcoming 
paper.

\section{Bremsstrahlung effects}\label{sec:bremss}
\begin{figure}[htb]
\centering
\includegraphics[height=7cm]{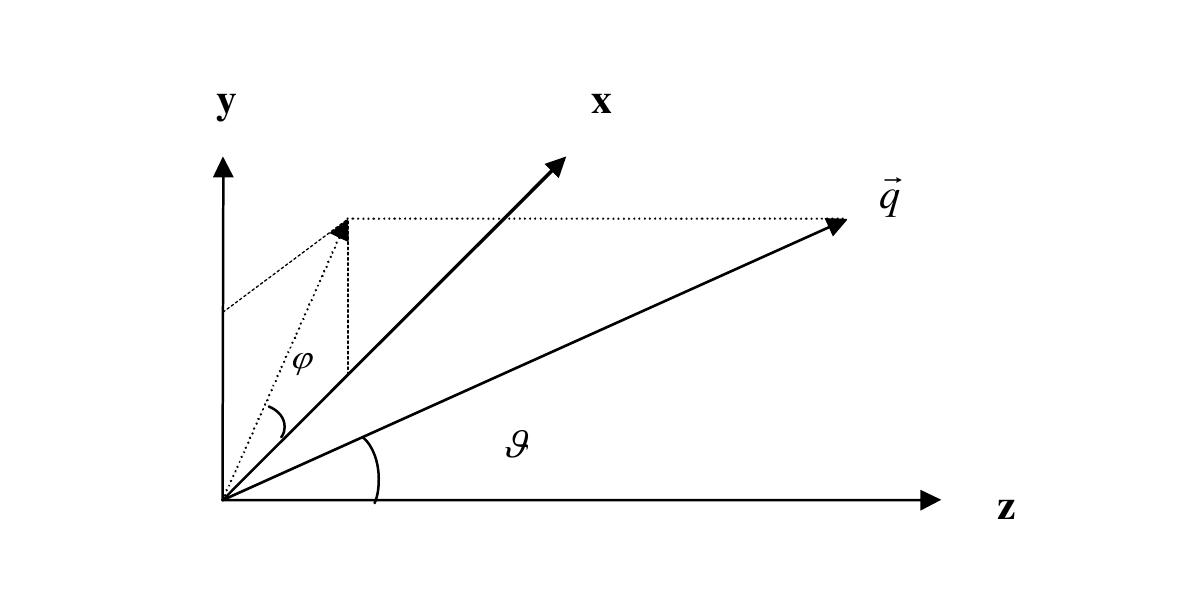}
\caption{ Reference frame definition for eq. (\ref{eq:Sbrem}). The magnetic field is directed along the $z$-axis.}
\label{fig:brenot} 
\end{figure}

The radiation emitted by an accelerated charge can be evaluated from the Lienard-Wiechert integral \cite{jacks} which yields the 
energy radiated per unit solid angle and unit frequency as
\begin{equation}\label{eq:liewie}
\frac{\mathrm{d}^2}{\mathrm{d} \Omega\,\mathrm{d} \omega}\,I \,=\, \frac{e^2 \omega^2}{4 \pi^2 c}\, S^2\;
\end{equation}
with $S$ modulus of the vector
\begin{equation}\label{eq:Sbrem}
\vec{S} \,=\, \int_0^T\,{\mathrm{d} t}\; \vec{q} \times (\vec{q} \times \vec{\beta})\, \exp \left\{\imath\,\omega 
\left(t - \frac{\vec{q} \cdot \vec{r}}{c}\right)\right\}\;,
\end{equation}
where $\vec{q}$ denotes the unit vector of the direction along which the emitted radiation is observed (see Fig. (\ref{fig:brenot})), 
$\vec{v} = \vec{\beta} c$ is the velocity of the charge, and $T$ is the time during which the charge effectively experiences 
the acceleration due to the fields.
By assuming that the motion occurs in absence of electric field and under the influence of a constant magnetic field, by using  
eq. (\ref{eq:posiz}) (with ($\vec{r}_0 = 0$) it is easy to show that:
\begin{eqnarray}\label{eq:expon}
\exp \left\{\imath\,\omega \left(t - \frac{\vec{q} \cdot \vec{r}}{c}\right)\right\} \!\!&=&\!\! \exp \left\{\imath \frac{\omega}{\Omega}\,c_1\right\}\, 
\exp \left\{\imath\,\omega\,c_2\,t\right\}\, \nonumber \\
&&\!\!\exp \left\{\imath \frac{\omega}{\Omega}\,\left[c_3\,\sin (\Omega t) \,-\, c_1\,\cos (\Omega t)\right]\right\}\;,
\end{eqnarray}
where ($\vec{n} = \vec{B}/B$, $\vec{\beta}_0 \,=\, \vec{v}_0/c$)
\begin{equation}
c_1 \,=\, \vec{q} \cdot (\vec{n} \times \vec{\beta}_0)\,,\quad
c_2 \,=\, \left[1 - (\vec{n} \cdot \vec{\beta}_0)\,(\vec{q} \cdot \vec{n})\right]\;, \quad
c_3 \,=\, (\vec{n} \cdot \vec{\beta}_0)\,(\vec{q} \cdot \vec{n}) \,-\, \vec{q} \cdot \vec{\beta}_0\;, \nonumber
\end{equation}
and, according to the Jacobi-Anger expansion \cite{andre}\footnote{The functions $B_k$ can be written in terms of the 
cylindrical Bessel functions by exploiting the identity $I_n (\imath \,x) = \imath^n\, J_n (x)$ and the Graf addition 
theorem \cite{andre}, which yields
\begin{equation}
B_n (x, \imath\,y) \,=\, \left(\frac{x + \imath\,y}{x - \imath\,y}\right)^{n/2}\,
J_n (\sqrt{x^2 + y^2})\;. \nonumber
\end{equation}
}
\begin{eqnarray}
\exp \left\{\imath \frac{\omega}{\Omega}\,\left[c_3\,\sin (\Omega t)\,-\,c_2\,\cos (\Omega t)\right]\right\} \!\!&=&\!\!
\sum_{k = - \infty}^\infty \,\mathrm{e}^{\imath k \Omega t}\,
B_k \left(\frac{\omega}{\Omega}\,c_3, - \imath\,\frac{\omega}{\Omega}\,c_2\right) \\
\!\!&=&\!\! \sum_{k = - \infty}^\infty \,\mathrm{e}^{\imath k \Omega t}\,\sum_{m = - \infty}^\infty\, 
J_{k - m} \left(\frac{\omega}{\Omega}\,c_3\right)\,I_{m} \left(- \imath\,\frac{\omega}{\Omega}\,c_2\right)\;. \nonumber
\end{eqnarray}
Moreover, it turns out
\begin{equation}
\vec{q} \times (\vec{q} \times \vec{\beta}) \,=\, \vec{a}_1 \,+\, \vec{b}_c\,\cos (\Omega\,t) \,+\, \vec{b}_s\,\sin (\Omega\,t)\;,
\end{equation}
where we introduced the following vectors
\begin{eqnarray}
\vec{a} \!\!&=&\!\! (\vec{n} \cdot \vec{\beta}_0)\,\left[(\vec{q} \cdot \vec{n})\,\vec{q} \,-\,\vec{n}\right]\;,  \nonumber \\
\vec{b}_c \!\!&=&\!\! (\vec{q} \cdot \vec{\beta}_0)\,\vec{q} \,-\, \vec{\beta}_0 \,-\, \vec{a}\,, \\
\vec{b}_s \!\!&=&\!\! \vec{q} \cdot (\vec{n} \times \vec{\beta}_0) \,-\, (\vec{n} \times \vec{\beta}_0)\;. \nonumber
\end{eqnarray}
Inserting these results in eq. (\ref{eq:liewie}), one obtains (for notation semplicity we have omitted to indicate the argument of the $B$-functions)
\begin{equation}\label{eq:skcomp} 
\vec{S} \,=\, \frac{T}{2}\,\exp \left\{\imath \frac{\omega}{\Omega}\,c_1\right\}\,\sum_{r = - \infty}^\infty \left[2\,\vec{a}\,B_{r} \,+\, 
\vec{b}\,B_{r - 1} \,+\, \vec{b}^\ast\,B_{r + 1}\right]\,\exp \left\{\imath \frac{\phi_r}{2}\,T\right\}\,
\mathrm{sinc}\left(\frac{\phi_r}{2}\,T\right)\;,
\end{equation}
where
\begin{eqnarray}
\vec{b} \,=\, \vec{b}_c \,+\, \imath\,\vec{b}_s\;, \qquad\qquad \phi_r \,=\, r\,\Omega \,+\, \omega\,c_2\;. \nonumber
\end{eqnarray}
According to the previous equation, the spectrum of $\vec{S}$ presents a series of harmonics with frequency
\begin{equation}
\omega_r \,=\, \frac{r\,\Omega}{1 - (\vec{q} \cdot \vec{n})\, (\vec{n} \cdot \vec{\beta}_0)}\;.
\end{equation}

In the case of a relativistic electron, we have
\begin{equation}
\Omega \,=\, \frac{e\,B}{\gamma\,m_0}\,=\, \frac{\omega_{c,0}}{\gamma}\;,
\end{equation}
and we can treat the solenoid as an undulator with period
\begin{equation}
\lambda_c \,=\, \gamma\,\lambda_{c,0} \,=\, \gamma\,\frac{2 \pi c}{\omega_{c,0}}\;.
\end{equation} 
If $\vec{n} = \vec{q}$ and the initial motion is mainly in the same direction of the field, we find 
\begin{equation}
\lambda_r \,\simeq\, \frac{\lambda_c}{2\,r\,\gamma^2}\;. 
\end{equation}

\begin{figure}[htb]
\centering
\includegraphics[height=9cm]{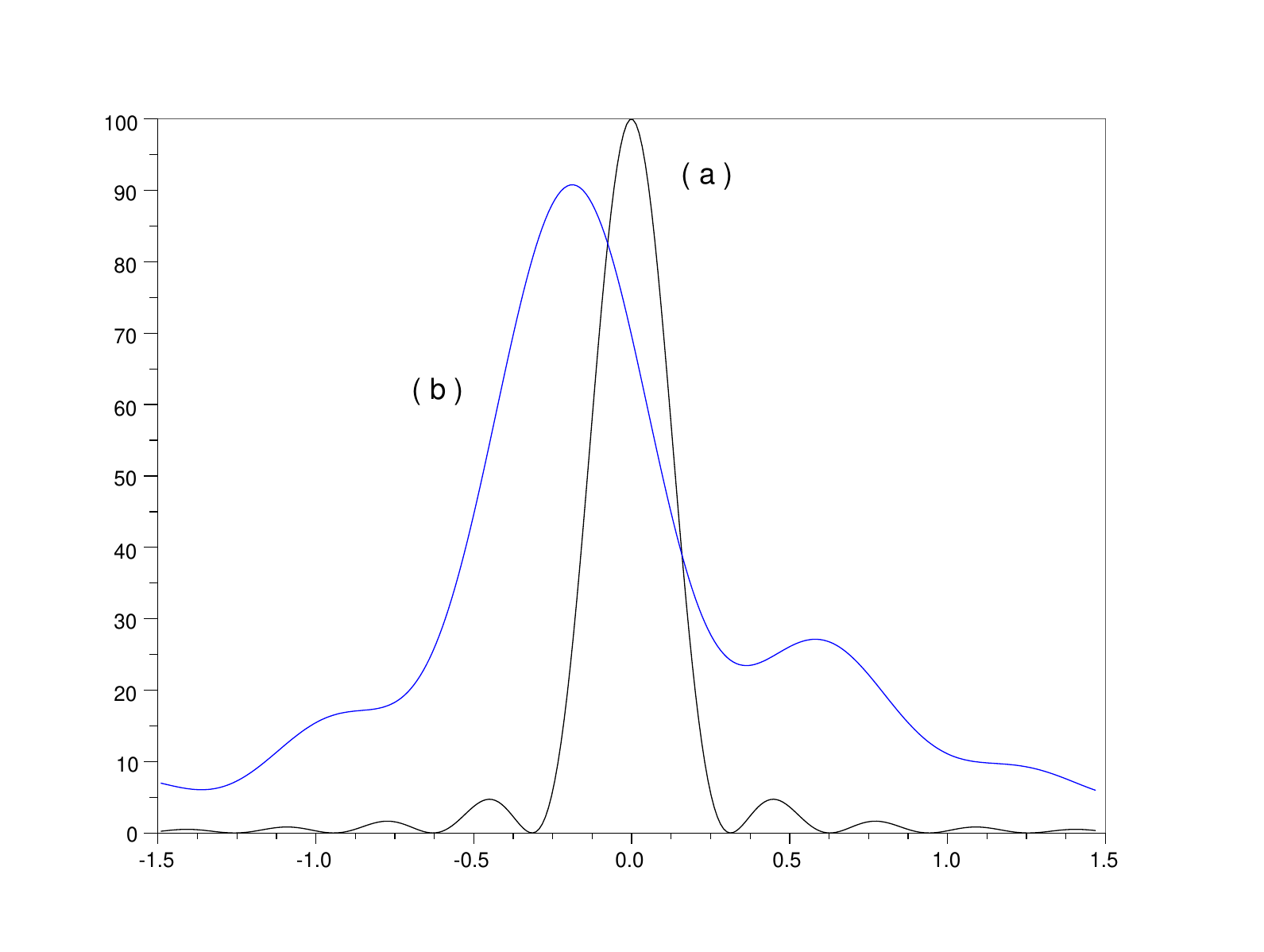}
\caption{Comparison of the emitted radiation lineshape between the case with only the 
solenoid (a) and the case where also an electric field is present (b).}
\label{fig:lines} 
\end{figure}

If we limit ourselves to the non relativistic case, the inclusion of the electric field does not imply any particular computational 
problem for the radiation integral. A significant difference is associated with the lineshape which is not simply the square of 
the sinc-function appearing in eq. (\ref{eq:skcomp}), but is now given by the square modulus of the function
\begin{equation}
F_r \,=\, \int_0^T\,\mathrm{d}t\, \exp\left(\phi_r\,t \,-\, \omega\,\frac{\vec{Q} \cdot \vec{q}}{c}\,t^2\right)\;.
\end{equation}
Fig. (\ref{fig:lines}) shows the shift and broadening of the spectral line of the radiation emitted by electrons moving in the field 
of a solenoid. 

\section{Quantum mechanical aspects and concluding remarks}\label{sec:quantum}
The final topic we will treat is the extension of these methods to quantum mechanics. To this aim we consider again the 
Hamiltonian given in eq. (\ref{eq:hamil}), but in absence of any electric field. If we assume that the magnetic field is directed along the 
$z$-axis, we can write the explicit form the Hamiltonian as follows\footnote{Note that in the static symmetric gauge 
(cfr eq. (\ref{eq:vspot})), $\vec{\nabla} \cdot \vec{A} = 0$.}
\begin{eqnarray}
\hat{H} \!\!&=&\!\! \frac{1}{2 m}\,\left(-\imath\,\hbar\,\vec{\nabla} \,-\, e\,\vec{A}\right)^2 \nonumber \\
\!\!&=&\!\! \frac{1}{2 m}\,\left\{- \hbar^2\,\nabla^2 \,+\, e^2\,B^2\,(x^2 + y^2) \,-\,\imath\,\hbar\,e\,B\,
(x \partial_y - y \partial_x)\right\}\;,
\end{eqnarray} 
which can be more conveniently cast in the form (see also \cite{JH} and references therein)
\begin{equation}
\hat{H} \,=\, - \frac{\hbar^2}{2 m}\,\partial_z^2 \,+\, \frac{1}{2 m}\,\omega_c^2\,r_L^2\,
\left(\hat{\Gamma}_1^2 + \hat{\Gamma}_2^2\right)\;
\end{equation}
where
\begin{equation}
\hat{\Gamma}_1 \,=\, \frac{1}{\sqrt{2}}\,(\imath\,\partial_\xi + \eta)\;\qquad 
\hat{\Gamma}_2 \,=\, \frac{1}{\sqrt{2}}\,(\imath\,\partial_\eta - \xi)\;\qquad
\xi \,=\, \frac{x}{x_L}\;\qquad \eta \,=\, \frac{y}{r_L}\;.
\end{equation}
The operators $\hat{\Gamma}_{1,2}$ satisfies the commutation rule
\begin{equation}
\left[\hat{\Gamma}_1,\,\hat{\Gamma}_2\right] \,=\, -\,\imath\;,
\end{equation}
and ($\hat{T} = \hat{\Gamma}_1 \hat{\Gamma}_2$)
\begin{equation}
\left[\hat{\Gamma}_1^2,\,\hat{\Gamma}_2^2\right] \,=\, - 4\,\imath\,\hat{T}\,+\,2\;,\qquad \qquad
\left[\hat{T},\,\hat{\Gamma}_k^2\right] \,=\,(-)^{k + 1}\,2\,\imath\,\hat{\Gamma}_k^2 \qquad (k = 1,2)\;,
\end{equation}
i.e. they exhibit the commutator properties of SU(1,1). The relevant Heisenberg equations write in the form of a vector 
torque equation of the same type of eq. (\ref{eq:cauchy}) and the time-dependent solution of the associated  Schr\"odinger 
problem can be obtained using the standard Wei-Norman ordering procedures \cite{weino}.

A different way of treating the quantum evolution problem consists in introducing a suitable transform as follows
($\vec{\zeta} = (\zeta_1, \zeta_2, \zeta_3)$) 
\begin{eqnarray}\label{eq:intrasf}
\hat{U} (t) \!\!&=&\!\! \exp\left\{- \frac{\imath t}{2 m \hbar}\,(\imath\,\hbar\,\vec{\nabla} + e\,\vec{A})^2\right\} \\
\!\!&=&\!\! \frac{1}{(2 \pi)^{3/2}}\,\int_{- \infty}^\infty\,\mathrm{d}\vec{\zeta}\,\mathrm{e}^{- \zeta^2}\,
\exp\left\{2\,\imath\,\sqrt{\frac{\imath t}{2 m \hbar}}\,\left[\vec{\zeta} \cdot (\imath\,\hbar\,\vec{\nabla} + e\,\vec{A})
\right]\right\}\;, \nonumber
\end{eqnarray}
where, from eq. (\ref{eq:vspot})
\begin{eqnarray}
 (\imath\,\hbar\,\vec{\nabla} + e\,\vec{A})_1 \!\!&=&\!\! \imath\,\hbar\,\partial_x \,-\,\frac{e}{2}\,B\,y\;, \nonumber \\
 (\imath\,\hbar\,\vec{\nabla} + e\,\vec{A})_2 \!\!&=&\!\! \imath\,\hbar\,\partial_y \,+\,\frac{e}{2}\,B\,x\;,\\
 (\imath\,\hbar\,\vec{\nabla} + e\,\vec{A})_3 \!\!&=&\!\! \imath\,\hbar\,\partial_z\;. \nonumber
\end{eqnarray}
In eq. (\ref{eq:intrasf}), by ordering the exponential in the integrand,  for the wave function at later time  we obtain
\begin{eqnarray}
\Psi (\vec{r}, t) \!\!&=&\!\! \hat{U} (t)\,\Psi (\vec{r})  \nonumber \\
\!\!&=&\!\! \frac{1}{(2 \pi)^{3/2}}\,\int_{- \infty}^\infty\,\mathrm{d}\vec{\zeta}\,\exp \left\{- \zeta^2 \,+\, 
\omega_c\,t\,\zeta_1\,\zeta_2 \,-\, \imath\,\sqrt{\frac{\imath\,m\,t}{2\,\hbar}}\,\omega_c\,(y\,\zeta_1\, - x\,\zeta_2)\right\} 
\nonumber \\
&& \qquad\qquad\qquad\qquad\qquad \Psi (\vec{r} - \sqrt{\frac{\imath\,2\,\hbar\,t}{m}}\,\vec{\zeta}\,)\;.
\end{eqnarray}

This relation is a generalization of the Gauss-Weierstrass transform and the study of its consequences will be discussed elsewhere.

It is worth stressing that the case where the direction of the magnetic field is arbitrary implies only a slightly more complicated integral 
transform. Even the inclusion of the electric field is not particularly tricky. Assuming that the electric field lies in the ($x, y$)-plane 
($\vec{E} = (E_x, E_y, 0)$), we can recast the Hamiltonian operator in the form
\begin{eqnarray}
\hat{H} \!\!&=&\!\! \frac{1}{2 m}\,\left(-\imath\,\hbar\,\vec{\nabla} \,-\, e\,\vec{A}\right)^2 \,-\, e\,\vec{E} \cdot \vec{r} \nonumber \\
\!\!&=&\!\!  \frac{1}{2 m}\,\left\{- \hbar^2\,\nabla^2 \,+\, \frac{e^2\,B^2}{4}\,\left[(x - \alpha_x)^2 \,+\, (y - \alpha_y)^2\right] \right. \nonumber \\ 
& &\qquad\qquad -\, \left. \imath\,\hbar\,\frac{e\,B}{2}\,\left[(x - \alpha_x)\,\partial_y \,-\, (y - \alpha_y)\,\partial_x\right]\right\} \nonumber \\
&& -\,\frac{1}{2}\,m\,\varpi^2\,(\alpha_x^2 +  \alpha_y^2) \,+\, \imath\,\hbar\,\frac{\varpi}{2}\,
(\alpha_x\,\partial_x + \alpha_y\,\partial_y)\;,
\end{eqnarray}
where
\begin{equation}
\alpha_{x,y} \,=\, \frac{4\,m\,E_{x,y}}{e B^2}\; \qquad\qquad \varpi \,=\, \frac{e\,B}{2\,m} \,=\,\frac{\omega_c}{2}\;. \nonumber
\end{equation}
This Hamiltonian is that of a multidimensional harmonic oscillator with shifted coordinates. The constant term 
$\frac{1}{2}\,m\,\varpi^2\,(\alpha_x^2 +  \alpha_y^2)$ is just the vacuum field energy redefinition associated with the transformation used 
to move to the shifted coordinate representation. The term proportional to $(\alpha_x\,\partial_x + \alpha_y\,\partial_y)$
is the quantum counterpart of the drift motion.

In the paper we have stressed the  analogy between Coriolis and Lorentz forces. This is more than a formal analogy and the 
previous considerations can be extended to the so called quantum Coriolis states \cite{daqua}, that are similar to the Landau quantum 
states \cite{landau} entering in the analysis of the motion of a quantum electron in a classical magnetic field. Some aspects of 
the problem and the possibility of studying them within the context of the present formalism, will be the argument of a forthcoming investigation.

The methods we have developed are flexible, fairly simple and easily amenable for numerical computation. Their use can also extended 
to non linear equations like the Landau-Lifshitz-Gilbert equation describing the precessional motion of the vector magnetization $\vec{M}$ in 
solids \cite{boldin}. Without entering in details, we remind that such equation, in our notation, writes 
\begin{equation}
\partial_t\,\vec{M} \,=\, - (\alpha + \beta\,\vec{M} \times)\,(\vec{M} \times \vec{H})\;.
\end{equation}
The quadratic non-linearity creates noticeable difficulties and the equation can be viewed as a kind of Riccati vector equation. 
Assuming however that the conditions of the problem allow the definition of a time step in which the variations of the vector $\vec{M}$ 
are not large, we can use the following solution scheme
\begin{equation}
\partial_t\,\vec{M}_n \,=\,\vec{P} \times \vec{M}_n\;, \qquad\qquad \vec{P} \,=\, (\alpha + \beta\,\vec{M}_{n -1} \times)\,\vec{H}\;,
\end{equation}
where the vector $\vec{P}$, containing the vector $\vec{M}$  at the previous integration step, is treated as a constant torque with the 
inclusion of an anti-damping term. The solution is essentially that provided in eq. (\ref{eq:corvel}), and the evolution should be followed by 
successive steps. 

The points in these concluding remarks have briefly treated just to show the flexibility of the method we have proposed and the number 
of topics which it allows to treat. Most of them deserves a deeper analysis, that we will develop in a forthcoming investigation.

\vspace{1in}
\noindent{\em Acknowledgments:} One of us (G. D.) expresses his sincere appreciation to the late Prof. 
J. O. Hirschfelder for deep discussions had almost twenty years ago about the contents of Ref. \cite{JH}, which 
stimulated the present paper.

\end{document}